\documentclass[twocolumn]{aastex631}

\newcommand{\qfit}{\texttt{q3dfit}}

\newcommand{\ly}{\hbox{Ly$\alpha$}}
\newcommand{\civ}{\hbox{C$\,${\scriptsize IV}}}
\newcommand{\heii}{\hbox{He$\,${\scriptsize II}}}
\newcommand{\oi}{\hbox{[O$\,${\scriptsize I}]}}
\newcommand{\oii}{\hbox{[O$\,${\scriptsize II}]}}
\newcommand{\oiii}{\hbox{[O$\,${\scriptsize III}]}}

\newcommand{\nii}{\hbox{[N$\,${\scriptsize II}]}}
\newcommand{\cii}{\hbox{[C$\,${\scriptsize II}]}}
\newcommand{\sii}{\hbox{[S$\,${\scriptsize II}]}}
\newcommand{\ha}{\hbox{H$\alpha$}}
\newcommand{\hb}{\hbox{H$\beta$}}

\newcommand{\niiha}{[N{\sc II}]/H$\alpha$}
\newcommand{\oiiihb}{[O{\sc III}]/\hb}
\newcommand{\siiha}{[S{\sc II}]/\ha}
\newcommand{\oiha}{[O{\sc I}]/\ha}

\newcommand{\kms}{km\,s$^{-1}$} 
\newcommand{\msun}{M$_{\odot}$} 
\newcommand{\eden}{cm$^{-3}$}

\newcommand{\ergs}{erg s$^{-1}$ }
\newcommand{\myr}{M$_\odot$~yr$^{-1}$} 
\newcommand{\loghn}{log(\nii/\ha) }
\newcommand{\logohb}{log(\oiii/\hb) }
\newcommand{\logsiiha}{log([S{\sc II}]/\ha)}
\newcommand{\logoiha}{log([O{\sc I}]/\ha)}
\renewcommand{\deg}{\ensuremath{^{\circ}}}
\newcommand{\ferg}{erg s$^{-1}$ cm$^{-2}$ }
\newcommand{\surff}{erg s$^{-1}$ cm$^{-2}$ \AA$^{-1}$arcsec$^{-2}$}

\newcommand{\lsun}{\ensuremath{\mathrm{L}_{\odot}}}

\shorttitle{Nuclear outflows and shocks in the CGM}
\shortauthors{Vayner et al.}

\graphicspath{{./}{figures/}}

\begin{document}

\title{Powerful nuclear outflows and circumgalactic medium shocks driven by the most luminous known obscured quasar in the Universe.}

\author[0000-0002-0710-3729]{Andrey Vayner}
\affiliation{IPAC, California Institute of Technology, 1200 E. California Boulevard, Pasadena, 91125, CA, USA}

\author[0000-0003-0699-6083]{Tanio Díaz-Santos}
\affiliation{Institute of Astrophysics, Foundation for Research and Technology–Hellas (FORTH), Heraklion, 70013, Greece}
\affiliation{School of Sciences, European University Cyprus, Diogenes street, Engomi, 1516 Nicosia, Cyprus}

\author{Peter R. M. Eisenhardt}
\affiliation{Jet Propulsion Laboratory, California Institute of Technology, 4800 Oak Grove Drive, Pasadena, 91109, CA, USA}

\author[0000-0003-2686-9241]{Daniel Stern}
\affiliation{Jet Propulsion Laboratory, California Institute of Technology, 4800 Oak Grove Drive, Pasadena, 91109, CA, USA}

\author[0000-0003-3498-2973]{Lee Armus}
\affiliation{IPAC, California Institute of Technology, 1200 E. California Boulevard, Pasadena, 91125, CA, USA}

\author[0000-0001-5769-4945]{Daniel Anglés-Alcázar}
\affiliation{Department of Physics, University of Connecticut, 196 Auditorium Road, U-3046, Storrs, 06269-304, CT, USA}

\author[0000-0002-9508-3667]{Roberto J. Assef}
\affiliation{Instituto de Estudios Astrof\'isicos, Facultad de Ingenier\'ia y Ciencias, Universidad Diego Portales, Av. Ej\'ercito Libertador 441, Santiago, Chile}

\author[0000-0002-7714-688X]{Román Fernández Aranda}
\affiliation{Institute of Astrophysics, Foundation for Research and Technology–Hellas (FORTH), Heraklion, 70013, Greece}
\affiliation{Department of Physics, University of Crete, Heraklion, 70013, Greece}

\author[0000-0001-7489-5167]{Andrew W. Blain}
\affiliation{School of Physics and Astronomy, University of Leicester, Leicester, LE1 7RH, UK}

\author[0000-0003-1470-5901]{Hyunsung D. Jun}
\affiliation{Department of Physics, Northwestern College,101 7th St SW, Orange City, 51041, IA USA}
\affiliation{School of Physics, Korea Institute for Advanced Study, 85 Hoegiro, Dongdaemun-gu, Seoul, 02455, Republic of Korea}

\author[0000-0002-9390-9672]{Chao-Wei Tsai}
\affiliation{National Astronomical Observatories, Chinese Academy of Sciences, 20A Datun Road, Beijing, 100101, China}
\affiliation{Institute for Frontiers in Astronomy and Astrophysics, Beijing Normal University, Beijing, 102206, China}

\author[0000-0002-0487-3090]{Niranjan Chandra Roy}
\affiliation{Department of Physics, University of Connecticut, 196 Auditorium Road, U-3046, Storrs, 06269-304, CT, USA}

\author{Drew Brisbin}
\affiliation{Instituto de Estudios Astrof\'isicos, Facultad de Ingenier\'ia y Ciencias, Universidad Diego Portales, Av. Ej\'ercito Libertador 441, Santiago, Chile}

\author[0000-0001-6266-0213]{Carl D. Ferkinhoff}
\affiliation{Winona State University, Winona, 55987, MN, USA}

\author[0000-0002-6290-3198]{Manuel Aravena}
\affiliation{Instituto de Estudios Astrof\'isicos, Facultad de Ingenier\'ia y Ciencias, Universidad Diego Portales, Av. Ej\'ercito Libertador 441, Santiago, Chile}

\author[0000-0003-3926-1411]{Jorge González-López}
\affiliation{Instituto de Astrof\'isica, Facultad de F\'isica, Pontiﬁcia Universidad Cat\'olica de Chile, Santiago 7820436, Chile}
\affiliation{Las Campanas Observatory, Carnegie Institution of Washington,  Ra\'ul Bitr\'an 1200, La Serena, Chile}

\author{Guodong Li}
\affiliation{National Astronomical Observatories, Chinese Academy of Sciences, 20A Datun Road, Beijing, 100101, China}
\affiliation{Instituto de Estudios Astrof\'isicos, Facultad de Ingenier\'ia y Ciencias, Universidad Diego Portales, Av. Ej\'ercito Libertador 441, Santiago, Chile}

\author{Mai Liao}
\affiliation{Instituto de Estudios Astrof\'isicos, Facultad de Ingenier\'ia y Ciencias, Universidad Diego Portales, Av. Ej\'ercito Libertador 441, Santiago, Chile}
\affiliation{National Astronomical Observatories, Chinese Academy of Sciences, 20A Datun Road, Chaoyang District, Beijing 100101, China}
\affiliation{Chinese Academy of Sciences South America Center for Astronomy, National Astronomical Observatories, CAS, Beijing, 100101, China}

\author[0000-0002-5033-8056]{Devika Shobhana}
\affiliation{Instituto de Estudios Astrof\'isicos, Facultad de Ingenier\'ia y Ciencias, Universidad Diego Portales, Av. Ej\'ercito Libertador 441, Santiago, Chile}

\author{Jingwen Wu}
\affiliation{National Astronomical Observatories, Chinese Academy of Sciences, 20A Datun Road, Beijing, 100101, China}
\affiliation{University of Chinese Academy of Sciences, Beijing, 100049, People's Republic of China}

\author[0000-0003-4293-7507]{Dejene Zewdie}
\affiliation{Centre for Space Research, North-West University, Potchefstroom, 2520, South Africa}

\correspondingauthor{Andrey Vayner}
\email{avayner@ipac.caltech.edu}

\begin{abstract}
We report integral field spectroscopy observations with the Near-Infrared Spectrograph on board \textit{JWST} targeting the 60 kpc environment surrounding the most luminous obscured quasar known at $z=4.6$. We detect ionized gas filaments on 40 kpc scales connecting a network of merging galaxies likely to form a cluster. We find regions of low ionization consistent with large-scale shock excitation surrounding the central dust-obscured quasar, out to distances nearly eight times the effective stellar radius of the quasar host galaxy. In the nuclear region, we find an ionized outflow driven by the quasar with velocities reaching 13,000 \kms, one of the fastest discovered to date with an outflow rate of 2000 \myr\ and a kinetic luminosity of 6$\times10^{46}$ \ergs\ resulting in coupling efficiency between the bolometric luminosity of the quasar and the outflow of 5\%. The kinetic luminosity of the outflow is sufficient to power the turbulent motion of the gas on galactic and circumgalactic scales and is likely the primary driver of the radiative shocks on interstellar medium and circumgalactic medium scales. This provides compelling evidence supporting long-standing theoretical predictions that powerful quasar outflows are a main driver in regulating the heating and accretion rate of gas onto massive central cluster galaxies.

\end{abstract}

\section{Introduction} \label{sec:intro}

Understanding how gas flows into galaxies and how it is regulated by feedback from accreting black holes and star formation is critical to understanding the growth of galaxies \citep{tuml11,kere05, keres09, vandeVoort11,Angles-Alcazar17,Angles-Alcazar21,Hopkins24}. The gas surrounding galaxies, or circumgalactic medium, encodes detailed information about these processes. It is extremely diffuse and historically has been studied in absorption along single sight lines to bright background sources \citep{zhu13, Prochaska14}. Two-dimensional studies have been typically restricted to rest-frame UV emission lines, particularly hydrogen \ly\ fluorescence, around luminous accreting supermassive black holes, i.e., quasars. Understanding the dynamics and ionizing source using the \ly\ line alone is challenging \citep{cant05,bori16,Cai19,Arrigoni-Battaia19,OSullivan20,Travascio20,Vayner23a,Sabhlok24}. The gas in the CGM is multi-phase with a range of temperatures and densities \citep{Zhu13b,Prochaska11b,Anderson16,Emonts18}, hosting a large fraction of baryonic matter \citep{Behroozi10,korm13}, with a significant amount of metals residing in the CGM \citep{tuml11,Tumlinson17,Rudie19}. Understanding the origin of gas turbulence and heating in the CGM \citep{tuml11,Rudie19,Rupke19} is critical to understanding how gas is supplied into the CGM and the subsequent feeding of the gas reservoir into galaxies for future star formation \citep{Angles-Alcazar17}.\\

The James Webb Space Telescope ({\it JWST}), in its first two years of science operations, has revolutionized our view of distant galaxies at near-infrared wavelengths \citep{Wylezalek22,Ding22,Finkelstein23,Veilleux23,Vayner23b,Rupke23,Vayner24}. Access to space-based near-infrared integral field spectroscopy (IFS) from 1-5 \micron\ \citep{Boker22,Rigby23} provides the necessary surface brightness sensitivity, wavelength coverage, and angular resolution to study in detail the internal dynamics, chemistry, and photoionization processes of galaxies, revealing the detailed physics of the feedback that regulates their growth at much earlier epochs than previously possible \citep{Parlanti24,DEugenio23,Perna23,Marshall23}. With {\it JWST}, we can now detect emission from optical recombination lines of hydrogen and collisionally excited metal lines on circumgalactic medium scales in the early Universe, when star-formation and quasar activity were vastly more intense \citep{Vayner23b}. Coupled with its high angular resolution, {\it JWST} can detect emissions originating from interactions between powerful quasar-driven outflows and the surrounding gas \citep{Vayner24,Chen25}. Here, we report on {\it JWST} detection of rest-frame optical emission from hydrogen, oxygen, sulfur, and nitrogen in the circumgalactic gas around the most luminous infrared galaxy known, which is seen 1.3 billion years after the Big Bang and hosts a dusty quasar. We map the emission out to distances beyond the stellar light extent of the quasar host galaxy using a mosaic observing strategy that covers a significantly larger field of view than previous spectroscopic studies with {\it JWST}, which were limited to galactic scales \citep{Vayner23b}.\\

We summarize target selection and observations in Section \ref{sec:obs} and data reduction in Section \ref{sec:data_reduction}. We present point-spread function subtraction, the construction of emission line maps, line ratios, and the investigation of the main source of gas ionization in Section \ref{sec:analysis}. We discuss the dynamics of the ionized gas and the driving source of the shocks in Section \ref{sec:dynamics}. We discuss our target in the context of other highly obscured quasars at high redshift in Section \ref{sec:disc}. We present our conclusions in Section \ref{sec:conc}. Throughout the article, we use a standard $\Lambda$CDM cosmology \citep{Planck13} with $\Omega_{M}$=0.308, $\Omega_{\Lambda}$=0.692, and H$_{o}$=67.8 \kms\ Mpc$^{-1}$. At $z=4.601$, 1\arcsec\ corresponds to 6.690 kpc, the age of the universe was 1.301 Gyr, and the look-back time is 12.496 Gyr. All magnitudes are on the AB scale unless otherwise stated. All wavelengths for emission lines quoted in the text are in vacuum wavelength, while the labels in the figures are often given with air wavelength.

\section{Target Selection and Observations}\label{sec:obs}

In cycle 1 with {\it JWST} we observed the most luminous (L$\rm_{bolometric}=~$3.5$\times10^{14}$\lsun) infrared galaxy in the observable universe \citep{Tsai18}, WISE J224607.57-052635.0 (W2246-0526) at $z=$ 4.601 (a look-back time of 12.5 Gyr). W2246-0526 is a hot dust-obscured galaxy (Hot DOG) selected by their red mid-IR colors in the WISE mission \citep{Wu12,Eisenhardt12}. The primary source of emission in W2246-0526 is from a dust-obscured quasar powered by accretion onto a supermassive black hole \citep{Tsai18,Fernandez-Aranda24}. The host galaxy of the luminous quasar is a compact massive galaxy with a stellar mass of 4$\times10^{11}$ \msun\ and an effective stellar light radius of 1.3 kpc \citep{diaz18,Fan18}. The galaxy is rich with molecular gas with a total molecular gas mass of 7$\times10^{10}$ \msun \citep{diaz18} and a total dust mass of 1.7$\times10^{8}$ \msun \citep{Fernndez-Aranda25}. W2246-0526 lives in an over-dense environment with many confirmed companion galaxies detected in \cii\ 158 \micron\ line with ALMA \citep{diaz18}, with many additional candidate Lyman-break galaxies detected in rest-frame UV imaging \citep{Zewdie23,Zewdie24}. The bolometric luminosity for W2246-0526 comes from integrating well-sample near-infrared to far-infrared spectral energy distribution \citep{Tsai18}. For hyper-luminous quasars, it is extremely challenging to measure their bolometric luminosities using known monochromatic bolometric correction factors as they only apply for quasars with luminosities in the range of $10^{45.1-47.3}$ erg/s range \citep{Runnoe12opt}. 

We map the host galaxy and the surrounding environment using the NIRSpec IFS by targeting the rest-frame optical emission lines of \oii $\lambda \lambda$ 3727.092, 3729.875, \hb, [OI] $\lambda$6300.304, \oiii $\lambda \lambda$4960.295, 5008.240, \nii $\lambda \lambda$6549.85, 6585.28, \ha, and \sii $\lambda \lambda$ 6718.29, 6732.67. 

NIRSpec observations using the integral field unit were taken on Jul 16, 2023 02:20:53 - 18:51:54 UT (ID 1712) using a combination of the G235H grating and F170LP filter and G395H grating and F290LP filter, resulting in wavelength coverage of 1.66$-$3.15 \micron\ and 2.87–5.14 \micron, respectively, with a spectral resolution of 85-150 \kms \citep{Jakobsen22}. We employed a 2$\times$3 mosaic observing strategy with a 4-point dither at each point within the mosaic. The detector was set up to readout in the NRSIRS2 mode with 15 groups per integration for an effective exposure time of 1094.167 seconds per dither per pointing for the G235H grating and 6 groups per integration for an effective exposure time of 452.25 seconds per dither per pointing for the G395H grating. We observed a leak-calibration exposure where the IFS aperture is closed at each point in the mosaic at the first dither position at each pointing with the same readout pattern. \\

\section{NIRSpec Data Reduction}
\label{sec:data_reduction}
Data reduction was made using the standard {\it JWST} pipeline (version 1.13.4), and Calibration Reference Data System (version 1193) with additional custom routines from \citet{Vayner24} based on previous experiences with NIRSpec IFS. We first run the ``Detector1Pipeline", which performs standard infrared array reduction steps such as ramp fitting, flagging cosmic rays, masking bad pixels, and dark and bias subtraction. We then correct the rate frames for 1/$f$ noise produced by the detector readout. We measure the 1/$f$ pattern per detector column using a running mean algorithm on rows where we do not expect any light from the multi-shutter array, fixed slits, or the IFS. The pattern is then subtracted on a column-by-column basis. We then run the ``Spec2pipeline" that assigns a world-coordinate system (WCS) to each frame, applies flat-field correction, wavelength calibration, and flux calibration, and produces 2D calibrated files. We then take a median of all the calibrated stage 2 files to search for static bad pixels that were not masked by the pipeline. These pixels are masked with not-a-number (\texttt{nan}) values and are properly flagged in the mask extension of each fits file. We run the ``cube build" algorithm on the bad pixel-masked frames and construct data cubes in IFS coordinates on a 50 mas grid using the ``drizzle" algorithm. We mask all sources in each cube at each dither position and smooth the cubes in the spectral direction using a running mean routine, and measure the background for each slice at each wavelength position using the ``Background2D" routine from the Photutils package \citep{larry_bradley_2023_7946442}. We then perform background subtraction on each dithered frame at each point in the mosaic to obtain a uniform background across the different spatial locations within the mosaic. We then align the cubes and project them onto a large mosaic frame using the flux-conserving reproject routine ``reproject exact" \citep{Robitaille20}. We use a sigma clip routine to determine outliers at each unique dither position and mask them prior to combining the data cubes. The final mosaic frame has an effective size of 9.5\arcsec $\times$ 6.7\arcsec. We combine the observations from the two gratings by projecting the cubes onto a common WCS grid and interpolating the spectra onto a common spectral grid. We achieve a final 2$\sigma$ flux sensitivity of 8.9$\times10^{-19}$ \surff\ and AB magnitude/arcsec$^{2}$ of 22.45 at 2.1 \micron, near redshifted \oii\ 3729 \AA; 8.6$\times10^{-19}$ \surff\ and AB magnitude/arcsec$^{2}$ of 22.45 at 2.79 \micron, near redshifted \oiii\ 5007 \AA; and 4.5$\times10^{-19}$ \surff\ and AB magnitude/arcsec$^{2}$ of 22.09 at 3.69 \micron, near redshifted \ha\ in a 0.4\arcsec$\times$0.4\arcsec\ aperture.

\section{\qfit\ analysis and results}
\label{sec:analysis}

\begin{figure}
    \centering
    \includegraphics[width=3. in]{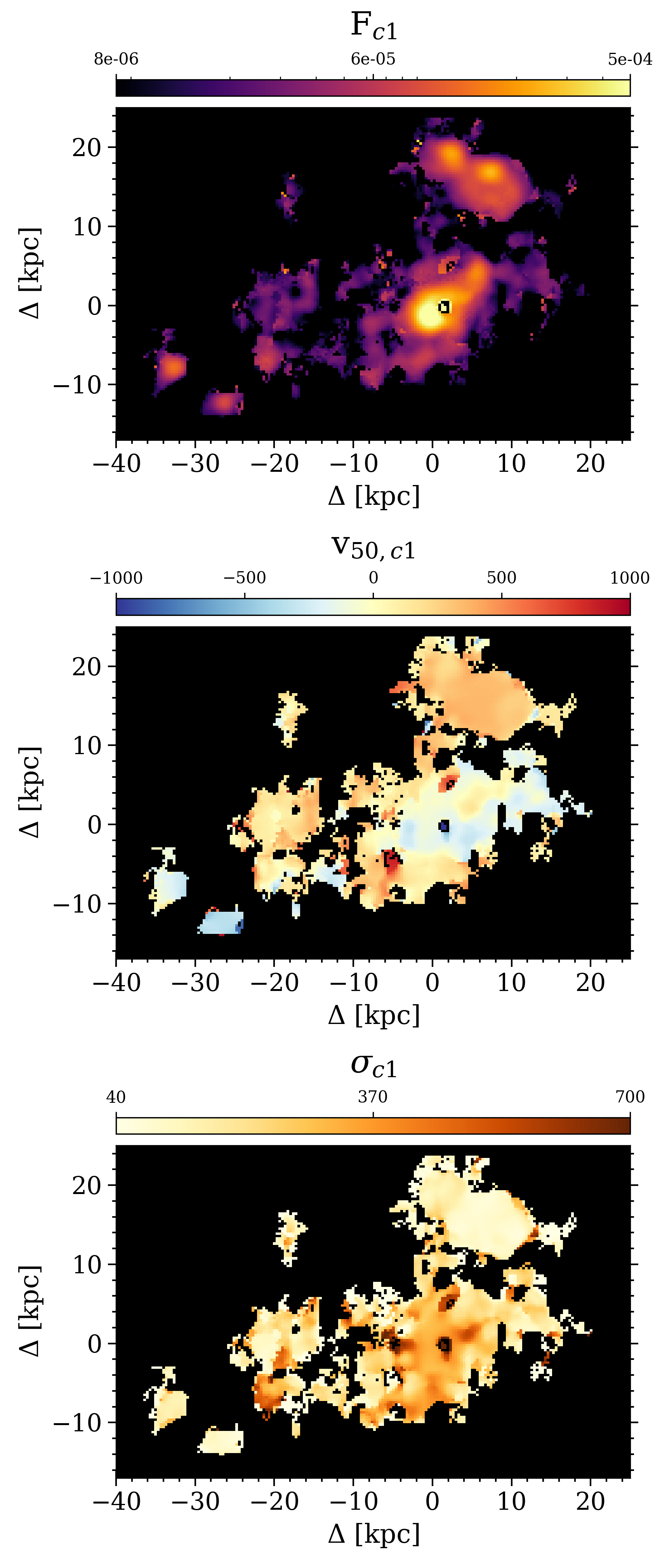}
    \caption{Extended emission in the W2246-0526 system from NIRSpec IFS observations. The top panel presents a map of the integrated intensity in the \ha\ emission line after PSF subtraction. The middle panel presents the \ha\ radial velocity map relative to the systemic redshift of the W2246-0526 system, and the bottom panel presents the \ha\ radial velocity dispersion map. The PSF subtracted quasar is located at position 0,0.}
    \label{fig:maps-ha}
\end{figure}

To subtract the unresolved emission from the quasar and to fit all extended emission lines and continuum in the mosaic, we use the \qfit\ software package \citep{q3dfit23} \footnote{\url{https://q3dfit.readthedocs.io/stable/}} designed to work with {\it JWST} NIRSpec and MIRI IFS. The software is based on the IFSFIT package \citep{rupk15} originally designed for ground-based IFS observations but was revamped to work with {\it JWST} in a user-friendly Python environment as part of an early release science {\it JWST} program (proposal ID 1128). The software package first produces a point-source dominated spectrum by integrating in a 1.5 pixel (0.15\arcsec) circular aperture centered on the point-source object. Data channels where we expect extended emission from specific lines are masked, and this spectrum is then fit as a ``QSO template" to the neighboring spaxels that are dominated by unresolved quasar emission. During the fitting, the template is scaled by a combination of polynomial and exponential functions to account for spectral and spatial variations of the PSF with an additional additive polynomial that accounts for spectral contribution of extended stellar emission from the host galaxy. The best fit ``QSO template" is subtracted out, and all emission lines above a specific threshold are fit with a Gaussian model. We fit \oii\ $\lambda$3727.092, 3729.875, \hb, [OI] $\lambda$6300.304, \oiii $\lambda$4960.295, 5008.240, \nii\ $\lambda$6549.85, 6585.28, \ha, and \sii\ $\lambda$6718.29, 6732.67 simultaneously, the redshift and line dispersion are all held fixed to the \ha\ line as it typically has the highest signal-to-noise ratio among the lines of interest. The model spectrum at each iteration in the least-squares fitting procedure is convolved with the line-spread function of the NIRSpec IFS at each wavelength. \oiii $\lambda$4960.30, 5008.24 lines are held fixed to a 1:2.98 \citep{Storey99} line ratio, \nii $\lambda$ 6549.85, 6585.28 are held fixed to 1:2.96 \citep{Galavis97} and the \oii $\lambda$ 3727.092, 3729.875 are held fixed to a 1:1.2 line ratio because the two lines are marginally spectrally resolved in some parts of the data, and completely unresolved in other regions. \sii\ $\lambda$6718.29, 6732.67 emission line ratios are allowed to vary between 0.5 and 1.5 based on the high and low-density limit ranges that these line ratios trace in $10^{4}$K gas \citep{Luridiana15}. We found that we required a single kinematic Gaussian component per emission line for most of the extended emission. The final output of \qfit\ are PSF-subtracted integrated intensity, velocity offset, and dispersion maps for each successfully fit emission line, a map of the extended continuum, and a PSF-subtracted data cube. We present the integrated intensity, radial velocity, and dispersion maps for the \ha\ line in Figure \ref{fig:maps-ha}. \\

\begin{figure*}
    \centering
    \includegraphics[width=5in]{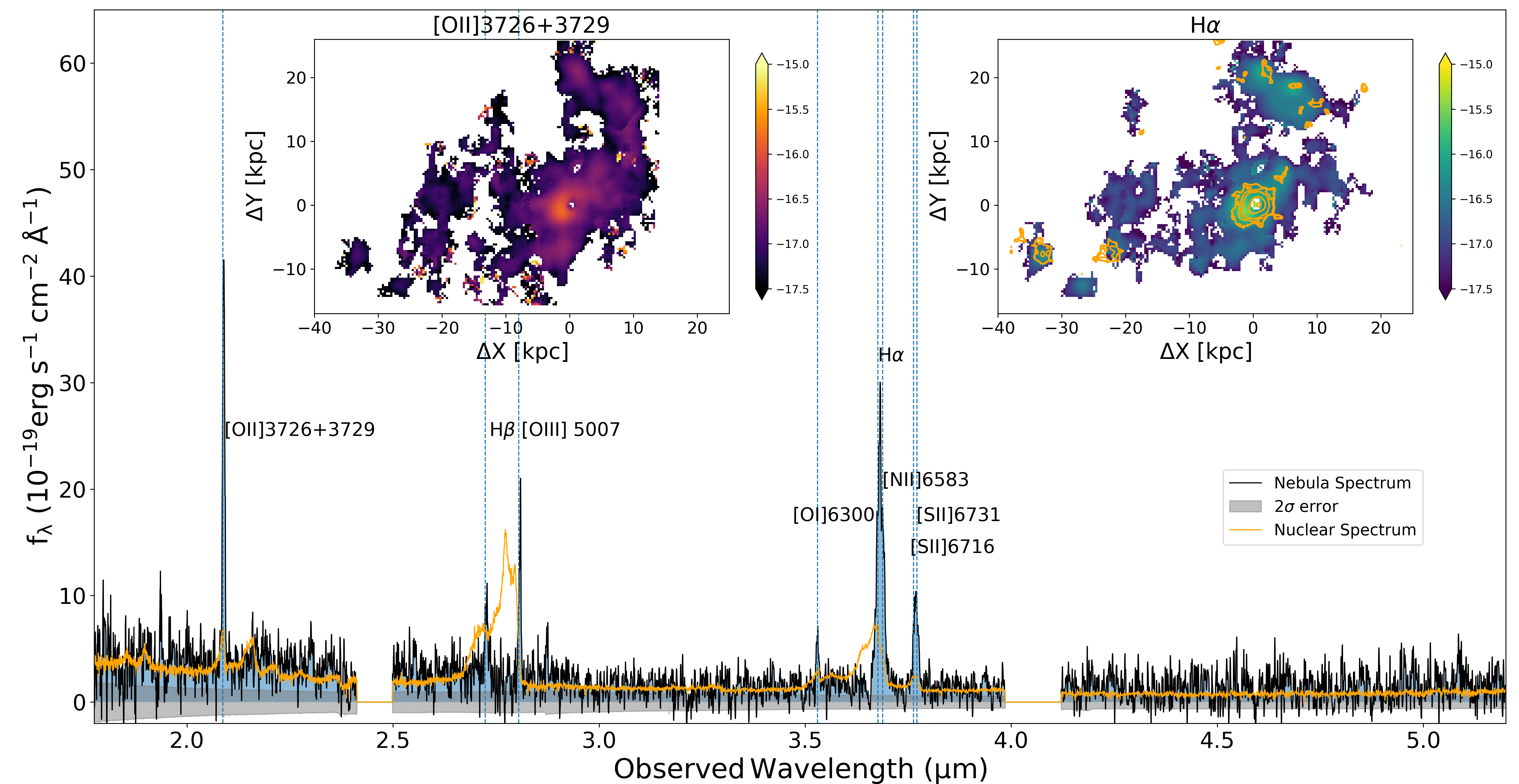}
    \caption{Near-infrared spectra of the W2246-0526 system at $z=4.601$. The black curve shows the observed 1.66-5.27\micron\ spectrum integrated over the entire nebula. The extended emission from the hot dust-obscured galaxy, the neighboring galaxies, and the extended circumgalactic medium are shown in the two inserted maps that trace the warm ionized gas through \oii\ and \ha\ emission. The spatially unresolved emission from the obscured quasar has been subtracted from these maps, and we construct the spectrum from a PSF subtracted data cube. Extended emission is detected out to 40 kpc from the central galaxy. The obscured quasar is located at position 0,0, and the color maps are shown in logarithmic units of \surff. We overlay the stellar continuum detected in the F160W filter with \textit{HST} in orange contours on the \ha\ map. The orange curve shows the spectrum of the spatially unresolved quasar emission ($r<1.2$ kpc) with strong blueshifted emission lines due to a quasar-driven outflow. The gray-shaded region shows the 2$\sigma$ uncertainty on the nebula integrated spectrum. }
    \label{fig:nebula_spec}
\end{figure*}

On a few kpc scales, we detect and resolve emission from the galaxy hosting the luminous obscured quasar and neighboring companion galaxies (Figure \ref{fig:nebula_spec}). On larger scales ($>10$ kpc), we detect the circumgalactic medium, which consists of clumpy emission and filaments that connect the central hot dust-obscured galaxy to its neighboring systems. On the integrated intensity \ha\ map, we overlay in contours the image of the stellar continuum from the companion galaxies detected with HST in the F160W filter (Figure \ref{fig:nebula_spec}). The filamentary structure extends towards two companion systems to the north and northwest and three companion galaxies towards the southeast. All targets are also detected in dust continuum \citep{diaz18} and in \cii\ 158 \micron\ emission along with the filaments that connect them \citep{diaz18}. All of the sources with detected stellar continuum have well-measured redshifts from both rest-frame optical lines in this study and far infrared observations with ALMA, and all fall within $\pm 700$\kms\ from the systemic redshift of W2246-0526. A future paper will include a detailed study of the companion galaxies and properties. We detect extended emission out to a maximum radial extent of 40 kpc from the quasar in the \oii\ and \ha\ emission lines, reaching surface brightness sensitivity of 5-10 \surff, consistent with the median radius of a typical \ly\ halo around a luminous quasar at this redshift \citep{Momose19,Drake19}. The search for extended \oii\ and \ha\ emission on circumgalactic medium scales at cosmic noon ($z\sim2$) and earlier epochs has proved challenging from the ground \citep{Langen23}. {\it JWST} observations suggest that when observed with sufficient surface brightness sensitivity, extended emission in these optical lines may be as common as \oii\ halos detected around lower redshift ($z<1.5$) quasars \citep{Johnson24}. The extended nebula observed in \oii\ and \ha\ shows a similar filamentary structure to what has been seen in low redshift group and cluster environments around luminous quasars \citep{Epinat18,Johnson18,Johnson22}. Similarly, in these lower redshift systems, the filamentary structure links nearby satellite galaxies to the quasar host galaxy. The size of the extended emission is consistent with the scale of the \cii\ and dust continuum emission, indicating the presence of a multi-phase circumgalactic medium consisting of both cold and warm gas on tens of kpc scales \citep{diaz18}. Dense, cool, and metal-rich circumgalactic medium gas might be common in quasar-host halos at this early epoch. \\

\subsection{Emission line ratios and ionization diagnostics}
\label{sec:line_ratio}

\begin{figure*}[!th]
    \centering
    \includegraphics[width=5 in]{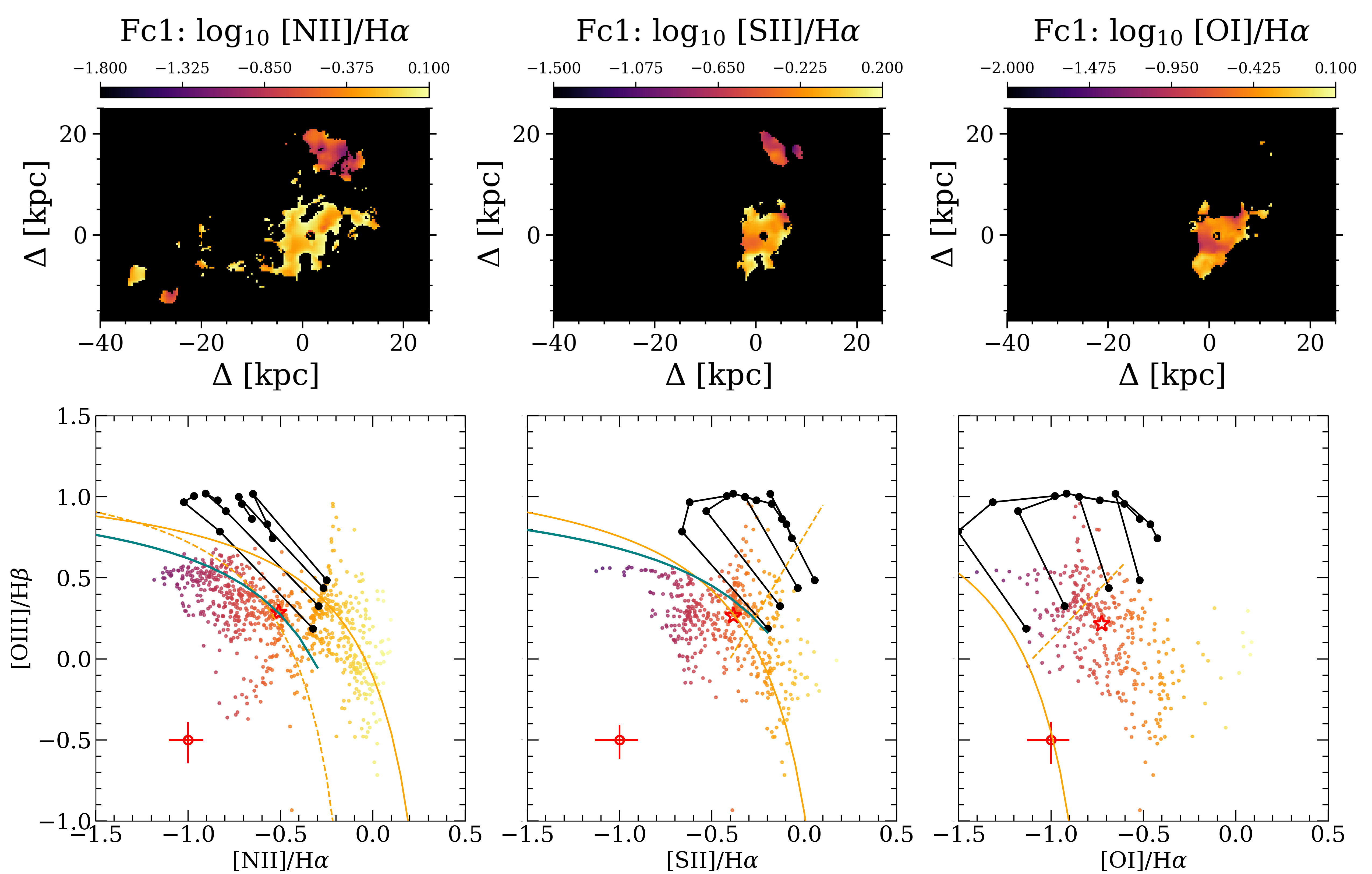}\\
    \caption{The top panel shows maps of emission line ratios between \niiha, \siiha\ and \oiha. The line ratio maps are constructed from PSF-subtracted data. The quasar is marked at position 0,0. The bottom panel shows the line ratios of \oiiihb\ plotted against \niiha, \siiha, and \oiha. The points are color-coded to their respective values on the x-axis, matched in color, and stretch to the top panel. The orange solid curve shows the maximum allowed theoretical line ratios for $z=0$ star-forming galaxies \citep{kewl01}, the dashed orange curve shows the empirical separation between star formation and AGN as source of ionization in SDSS galaxies \citep{kauf03a}, while the teal line shows the locus of the star formation metallicity sequence at $z=2-3$ \citep{Strom17}. We present the quasar photoionization model tracks for solar metallicity gas in black \citep{grov04a}. The red circle point shows the median uncertainty on the line ratios, while the red star shows the median line ratio.}
    \label{fig:line-ratio-BPT_maps}
\end{figure*}

\begin{figure*}[!th]
    \centering
    \includegraphics[width=5 in]{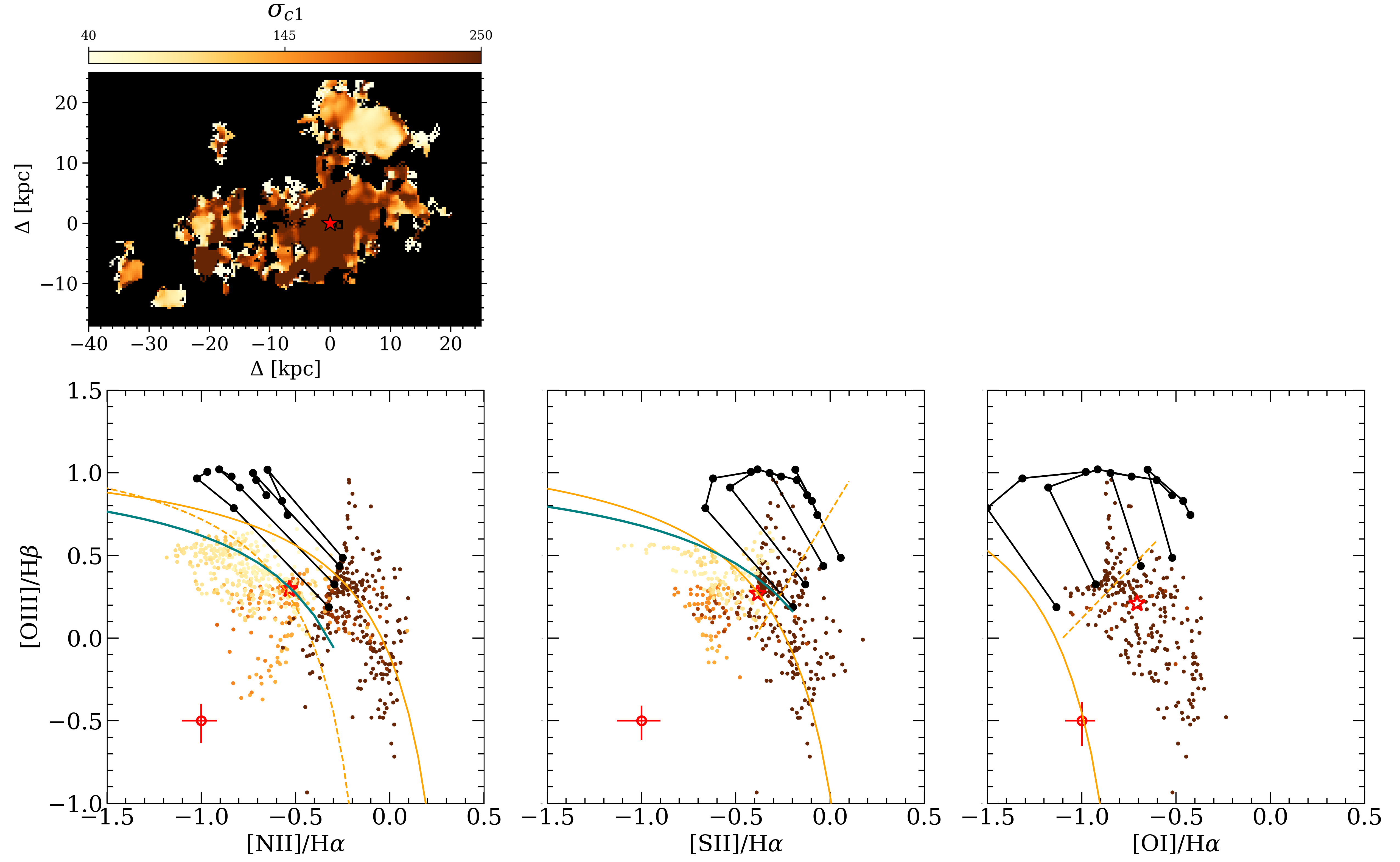}
    \caption{Similar figure to \ref{fig:line-ratio-BPT_maps}, however, the points on the line ratio plots (bottom panel) are color-coded to the velocity dispersion values (top panel, similar to Figure \ref{fig:maps-ha}), with dark brown points representing gas with $v_{\sigma}>250$ \kms\ and lighter orange points representing gas with $v_{\sigma}<250$ \kms. On average, high-velocity dispersion regions are consistent with the lowest ionization points. The high and low-velocity dispersion points clearly showcase different ranges of emission line ratios consistent with different sources of ionization. The red circle point shows the median uncertainty on the line ratios, while the red star shows the median line ratio.}
    \label{fig:line-ratio-BPT}
\end{figure*}

To investigate the source of ionization across the W2246-0526 system, including its merging galaxies and extended circumgalactic medium, we construct maps of \niiha, \siiha, and \oiha\ (top panel Figure \ref{fig:line-ratio-BPT_maps}). We only include spaxels where both lines are detected with a $3\sigma$ or higher significance. We plot \oiiihb\ vs. \niiha, \siiha, and \oiha\ line ratios and compare them to models of various sources of ionization (Figure \ref{fig:line-ratio-BPT}). We overplot dusty quasar photoionization models for solar metallicity gas on all emission line ratio diagrams \citep{grov04a}. For a given metallicity, the line ratios vary based on changes in the ionization parameter (log($U$): [-2,0]) and the slope of the ionizing power-law spectrum ($\alpha$: [-2,-1.2]). Extended narrow line regions of $z\sim2$ quasars often require ionization parameters and ionizing power-law spectrum slopes in our selected range \citep{Vayner21_ion}. Our choice for selecting solar metallicity models is motivated by the fact that ALMA observations of W2246-0526 far-infrared fine-structure lines are most consistent with 0.5-1 $Z_{\odot}$ \citep{Fernandez-Aranda24}. Dust plays an important role in radiation pressure confinement \citep{Stern14}. On kpc scales based on the \ha\ and \hb\ line ratios and assuming case B recombination, we estimate extinction values in the V band reaching 2 magnitudes. This indicates that there is a plethora of dust on kpc scales. Hence, the dusty radiation pressure-dominated models from \citet{grov04a} are the most appropriate to compare to. There is a lack of points overlapping with the quasar photoionization models, which indicates that the central powerful quasar is not photoionizing a significant portion of the extended emission. Quasar photoionization models at lower metallicity shift to lower \niiha\ values, making them less consistent with our observed line ratios on the \oiiihb\ vs. \niiha\ diagram.\\

\begin{figure}
    \centering
    \includegraphics[width=3.4in]{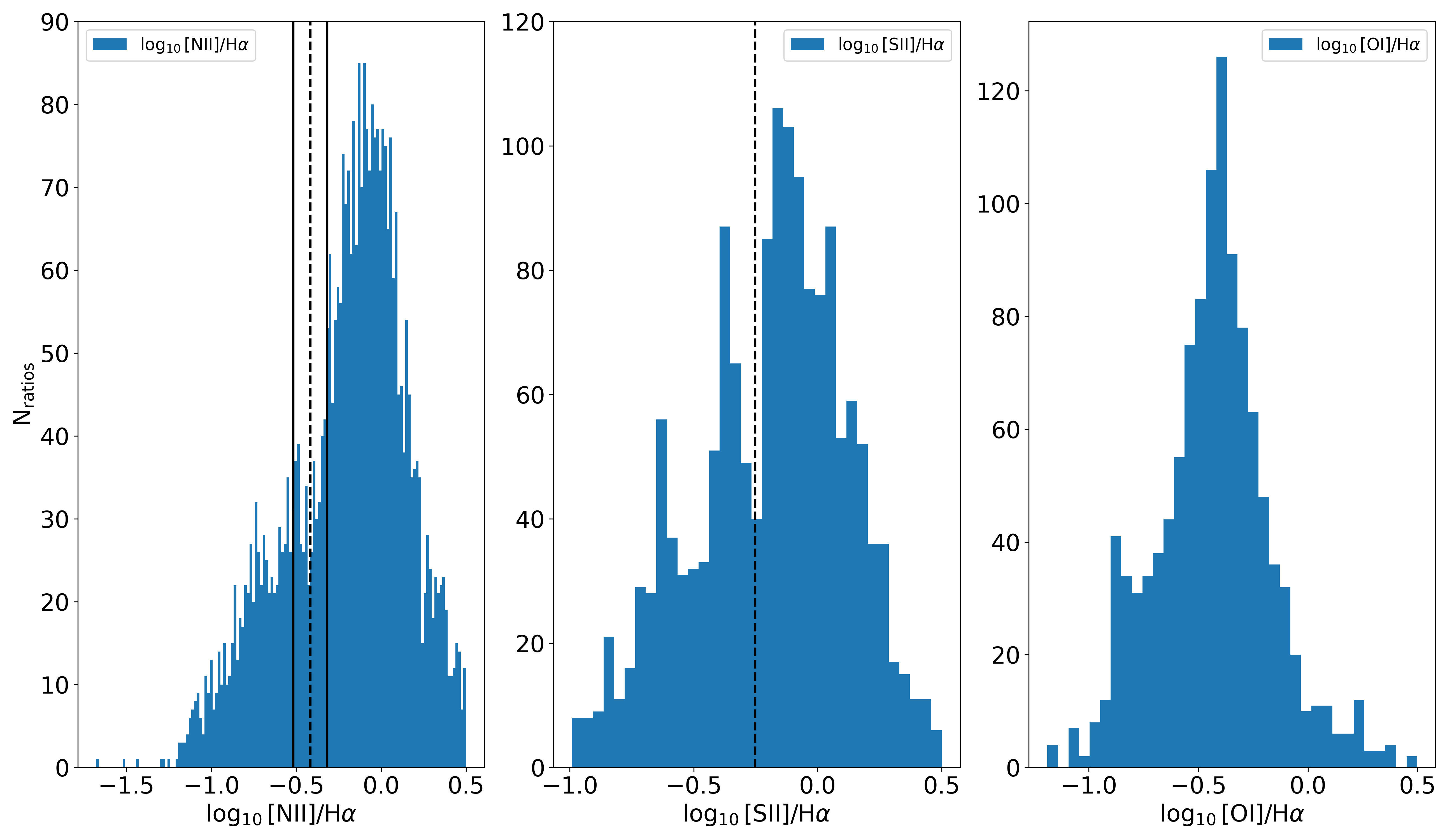}
    \caption{Histograms showing line ratio distribution of \niiha, \siiha\ and \oiha. We see a broad distribution in the \niiha\ and \siiha\ histograms, where the higher line ratios correlate with higher velocity dispersion. The thick black lines show the range of points that may be contaminated by ionization from star formation from the double distribution of emission line ratios and are used to estimate an upper limit on the recent unobscured star formation rate. We see a single distribution in the \oiha\ line ratio because \oi\ is only detected in a nebula centered around the quasar where the line ratios are elevated.}
    \label{fig:hist-ratios}
\end{figure}

\begin{figure}
    \centering
    \includegraphics[width=3.0in]{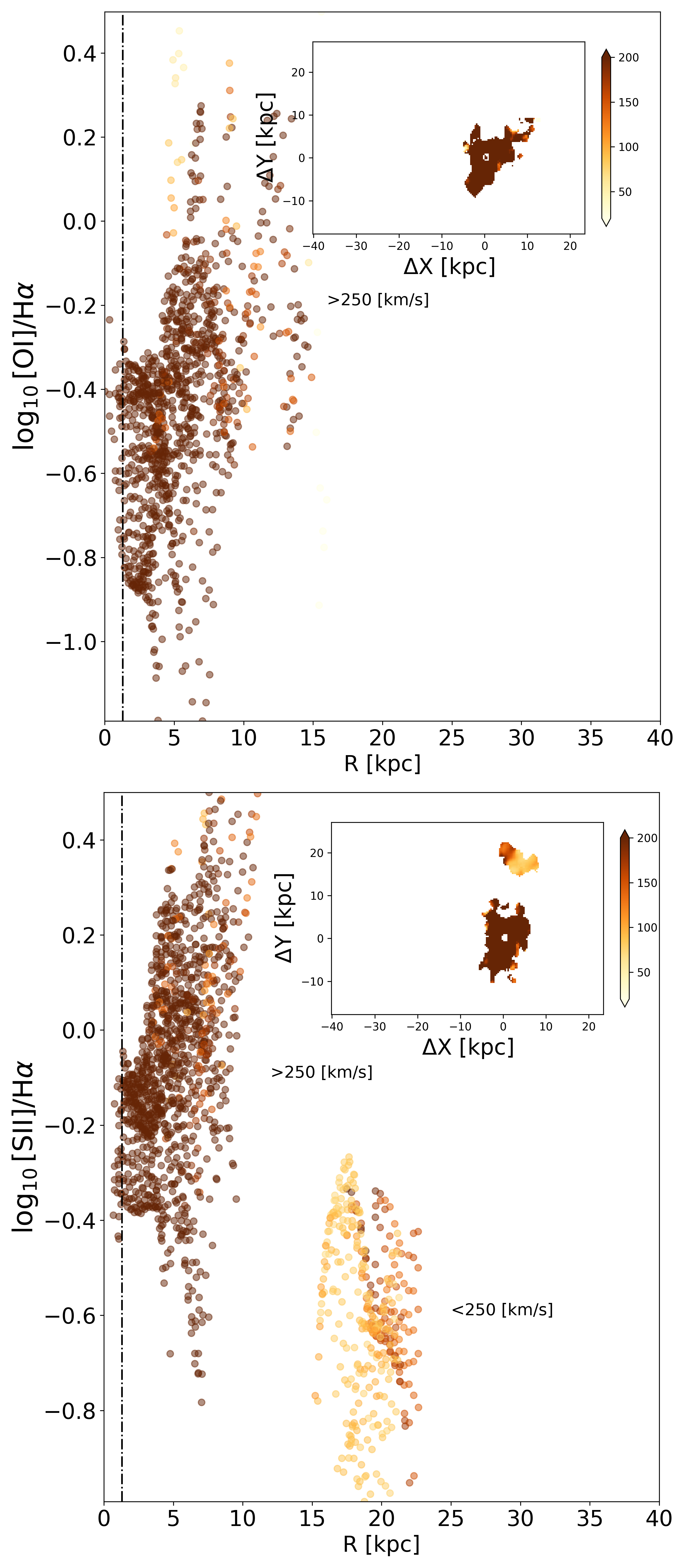}
    \caption{Line ratio of \oiha\ and \siiha\ plotted against the projected separation from the quasar. The points are color-coded to their velocity dispersion value. High-velocity dispersion values are shown in dark brown, while the low-velocity dispersion points are shown in orange. High line ratios with high-velocity dispersion values are found in a 10 kpc radius around the quasar, consistent with shock heating as the source of the elevated line ratios. Lower line ratios are predominantly located at distances $>15$ kpc in the companion galaxies and their extended interstellar and circumgalactic medium. These line ratios are consistent with photoionization by star formation. The \oi\ line is only detected in the 10 kpc radius nebula centered on the quasar. We present maps of velocity dispersion where we detect \oi\ and \sii\ in each line ratio vs. distance panel, on a hard stretch to emphasize the regions of low and high velocity dispersion.}
    \label{fig:ratio_vs_distance}
\end{figure}

We see that the majority of the points lie above the line separating star formation from quasar ionization based on the local SDSS galaxy sample. Yet, the line ratios are inconsistent with quasar photoionization due to lower \oiiihb\ line ratios indicating a different source of gas ionization than star formation or a typical luminous quasar. There are two distinct populations of emission line ratios, one with \loghn $<-0.3$ and a second population with \loghn $>-0.3$ (Figure \ref{fig:hist-ratios}). The lower \niiha\ line ratios are associated with companion galaxies and their surrounding nebula. They are consistent with ionization by young massive stars, showing a clear star formation-metallicity sequence on the BPT diagram \citep{Kewley13b}. We overplot the locus of the star formation-metallicity sequence (teal line, Figure \ref{fig:line-ratio-BPT_maps}) from observations of star-forming galaxies in the stellar mass range $10^{9-11.5}$ \msun\ at $z=2-3$ with star formation rates ranging from 3-1000 \myr\ \citep{Strom17}. The higher \niiha\ ratios are centered around the quasar in a 10 kpc scale nebula and correlate with regions of the highest velocity dispersion (bottom panels in Figure \ref{fig:line-ratio-BPT} and Figure \ref{fig:ratio_vs_distance}). The regions of high \niiha\ ratios are also where we see higher \siiha\ and \oiha\ ratios that also correlate with high-velocity dispersion ($V_{\sigma}>$ 250 \kms) and often lie outside the maximum allowed line ratios by O and B star ionization in the \oiiihb\ vs. \siiha\ and \oiha\ diagrams. The \niiha\ line ratios are also beyond what is observed in star-forming galaxies at cosmic noon \citep{Strom17}. Furthermore, we see that points with the highest \niiha, \siiha, and \oiha\ are consistent with the lowest \oiiihb\ ratios, indicating the presence of low-ionization gas. 

\begin{figure*}[!th]
    \centering
    \includegraphics[width=5in]{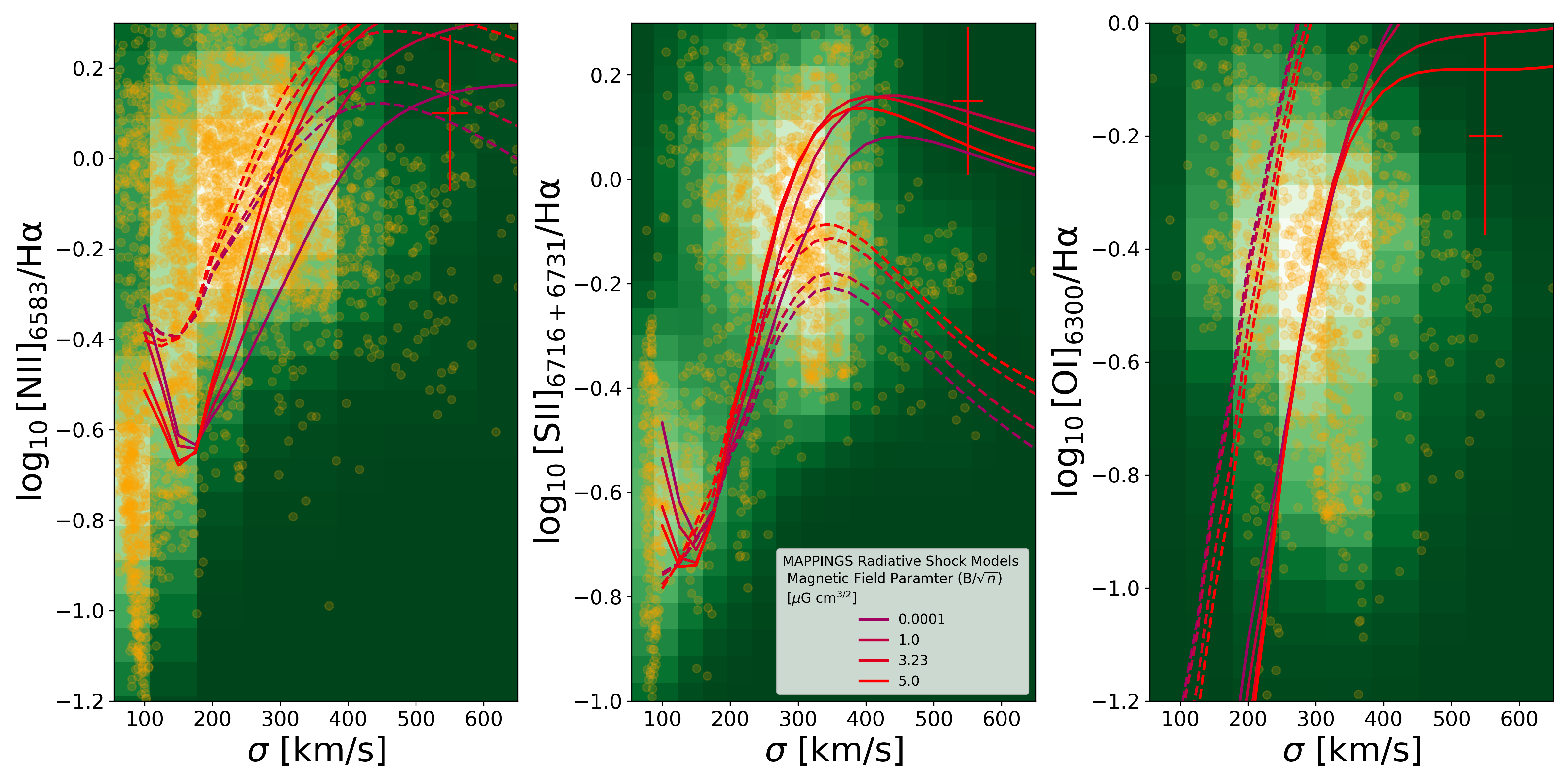}\\
    \includegraphics[width=5in]{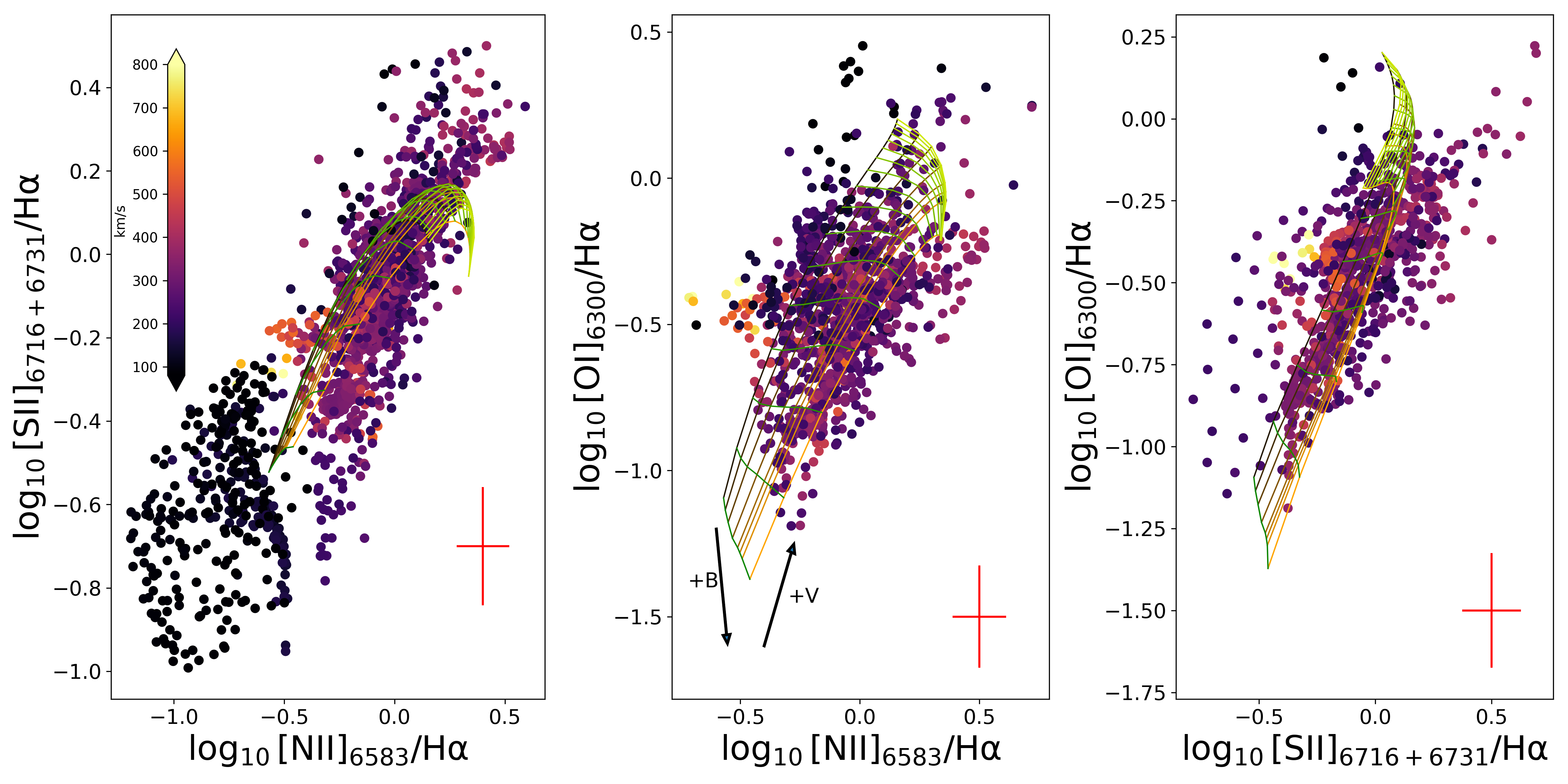}
    \caption{Top: \niiha, \siiha\ and \oiha\ line ratios vs. velocity dispersion diagrams. The orange symbols show the measured values at each spaxel while the shading of the rectangles indicates the number density of measured values with that line ratio and velocity. The median uncertainties are shown with red crosses. The red-purple curves show MAPPINGS radiative shock models for a range of magnetic field parameters for gas with pre-shock electron density of 10 \eden\ (thick line) and 100 \eden\ (dashed line). In all diagrams, the highest observed line ratios at the largest velocities are consistent with the expected line ratios due to shocks. While \oi\ is only detected at high-velocity dispersions ($V_{\sigma}>200$ \kms), \nii\ and \sii\ are observed across a range of galactic and circumgalactic regions over a range of velocity dispersions, with the \niiha\ and \siiha\ line ratios increasing with velocity dispersion and a hint for a subsequent decrease at the highest velocities, as predicted by radiative shock models. Bottom: line ratios of \siiha, \niiha\ and \oiha\ plotted against each other. The points are color-coded to the velocity dispersion, with purple to orange points representing the high-velocity dispersion ($V_{\sigma}>200$ \kms) and black points with low-velocity dispersion ($V_{\sigma}<200$\kms). MAPPINGS shock models are overplotted and are able to explain the line ratios across all ions and neutral oxygen.}
    \label{fig:ratio_vs_dispersion}
\end{figure*}

The emission line ratios for the high-velocity dispersion gas are consistent with low ionization; \logohb\ $<0.5$, \loghn\ $>-0.3$, \logoiha\ $>-1$, and \logsiiha $>-0.3$ (see Figure \ref{fig:ratio_vs_dispersion}). The highest velocity dispersion gas shows the highest observed \niiha, \oiha, and \siiha\ line ratios across the inner 10 kpc nebula as well as compared to the companion galaxies. In the nearby universe, the \nii, \oi\ and \sii\ lines have been used as tracers of shock-ionized gas \citep{rich10,rich11} and within planetary nebulae in our own Milky Way galaxy \citep{Phillips98}. Radiative shocks in the interstellar medium are used to explain the enhancement of \niiha, \siiha\ and \oiha\ line ratios at higher velocity dispersions \citep{Phillips98,rich10,rich11,Vayner24,Wang24,Lamperti24}.  

\begin{figure*}
    \centering
    \includegraphics[width=5in]{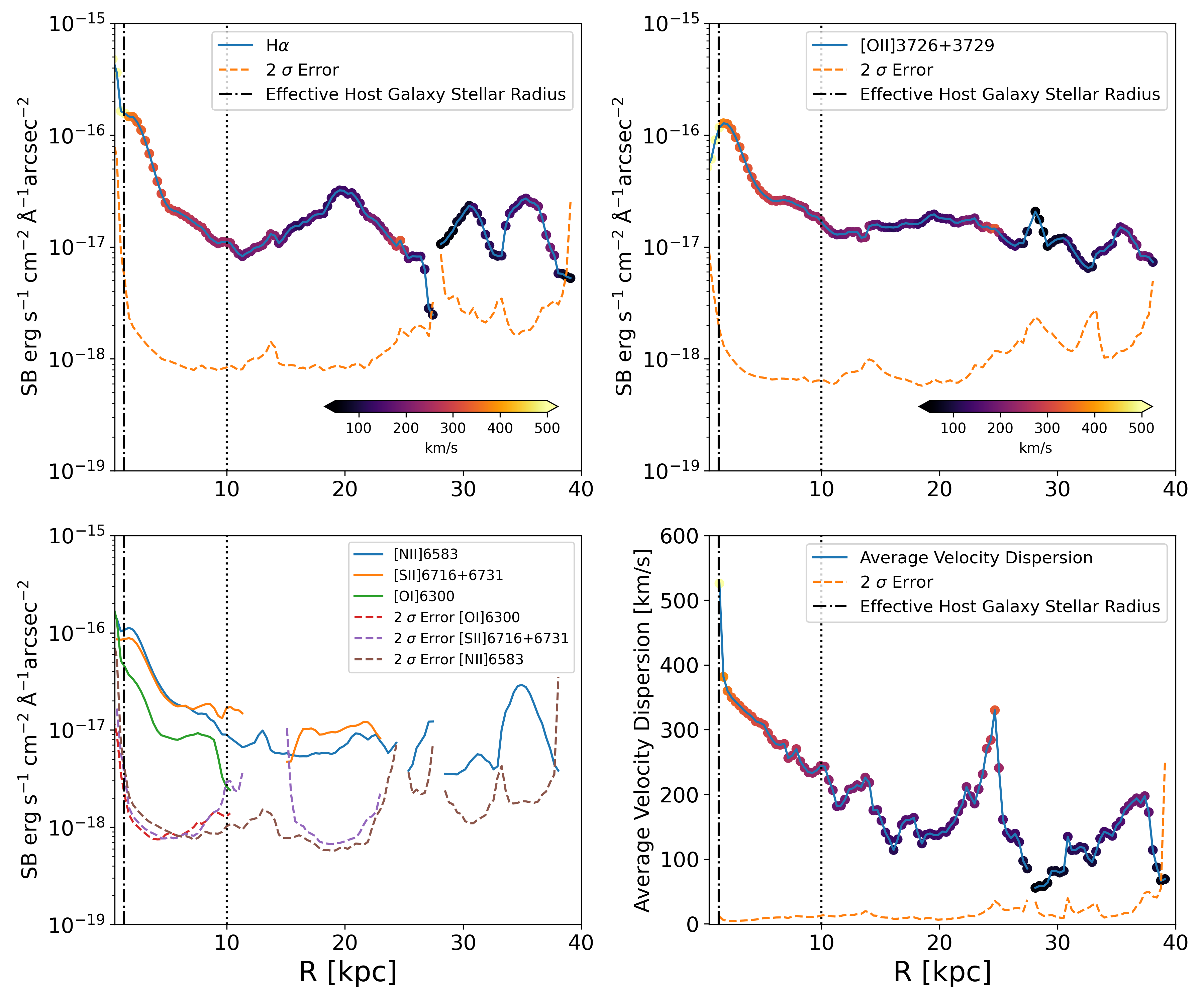}
    \caption{Radial profiles of the ionized gas for the extended emission in the W2246-0526 system. The top left plot shows the \ha\ surface brightness profile, computed in concentric rings of 0.05\arcsec\ width centered on the quasar in the PSF subtracted data. The points are color-coded to the average velocity dispersion measured in the concentric ring. The orange dashed line shows the uncertainty on the surface brightness profile. The top right plot shows a similar diagram for \oii\, and the bottom left plot shows the surface brightness profile for fainter lines of \nii, \sii, and \oi. The bottom right plot shows the average velocity dispersion measured in each concentric ring. We see a decline in the average velocity dispersion with a transitional radius from higher ($V_{\sigma}>$250\kms) to lower velocity dispersion around 10 kpc, which is highlighted in each diagram as a dotted vertical line. Line ratios interior to this radius are most consistent with shock excitation. The dot-dash vertical line shows the stellar effective radius of the hot dust-obscured galaxy of 1.3 kpc \citep{Fan18}. High-velocity dispersion gas that is consistent with low-ionization is detected out to a radius of at least 10 kpc, about 8 times the radius of the stellar extent, well into the circumgalactic medium scale.}
    \label{fig:SB-plot}
\end{figure*}

In Figure \ref{fig:ratio_vs_dispersion}, we plot the \niiha, \siiha\ and \oiha\ emission line ratios against velocity dispersion and compare them to predicted line ratios based on radiative shocks from MAPPINGS \citep{alle08} at a range of shock velocities. We specifically select models from the library of shock models from the 3MdB database \citep{Alarie19}. These models consist of pure shock models, pure precursor, and shock and precursor models. We use the models computed using the chemical abundances from \citep{Gutkin16}. We explore all available models at a range of pre-shock density, magnetic field parameter (B/$\sqrt{n}$), shock velocities, and the full range of available abundances.  We see that the elevated line ratios at the highest velocity dispersion can be explained by pure radiative shocks where the ISM has a pre-shock electron density of 10 \eden\ and a magnetic field parameter (B/$\sqrt{n}$) of 0.0001-10 $\mu$G cm$^{3/2}$ using the \texttt{Gutkin16\_ISM0d030\_C1d00} abundance set. The most likely scenario is that shocks ionize the gas in a 10 kpc radius around the quasar. The high-velocity dispersion, low-ionization gas extends up to 10 kpc $\sim8$ times the effective radius of the stellar light \citep{Fan18} (Figure \ref{fig:SB-plot}), indicating the heating of gas on circumgalactic medium scales through radiative shocks, providing strong evidence for mechanical feedback on circumgalactic scales.  

\section{Dynamics of the ionized ISM/CGM}
\label{sec:dynamics}

To understand the driving mechanism behind large-scale shocks in W2246-0526, we need to estimate the energetics of the shocked gas and compare it to the available input energy from various feedback sources. By assuming that the majority of the \ha\ emission over the high-velocity dispersion and low ionization region is dominated by radiative shocks, we can estimate the total kinetic energy due to turbulence in the interstellar and circumgalactic medium in the warm ionized gas phase. First, we correct the integrated intensity \ha\ map for dust reddening using the Balmer decrement method and the \citet{Calzetti2000} extinction law, assuming any deviation from the case B recombination (\ha/\hb $=$ 2.86) line ratio is due to dust reddening. We measure a total \ha\ luminosity of $1\times10^{44}$ \ergs\ by summing the flux in the spaxels that show $v_{\sigma}>250$ \kms. Assuming case B recombination, solar abundance for He, and constant electron density across the nebula where we detect ionized emission, where each cloud has the same density (fill factor of unity), we can convert the \ha\ luminosity into an ionized gas mass using the following equation \citep{oste06}:

\begin{equation}\label{equation:ionized_gas_mass}
    M_{ionized} = 1.4 \bigg(\frac{m_{p}L_{H\alpha}}{j_{H_{\alpha}}n_{e}}\bigg)
\end{equation}

\noindent where $\rm L_{H\alpha}$ and $\rm n_{e}$ are \ha\ luminosity and the average electron density over the shock region. For the electron temperature we assume a range of (1-2)$\times10^{4}$ K constraining the \ha\ line emissivity to (1.8-3.53)$\times10^{-25}\rm~erg~cm^{3}~s^{-1}$ for an electron density of $\sim10^{2-3}$ \eden \citep{Luridiana15}. 

\pagebreak

We estimate the total kinetic energy due to turbulent motion to be $3\times10^{58}$ erg using the following equation:

\begin{equation}
   \rm E_{turb} = \frac{3}{2} M_{ionized}\times\sigma_{m}^{2}
    \label{eq:turb}
\end{equation}

\noindent for a warm ionized gas mass of 6$\times10^{9}$\msun (Equation \ref{equation:ionized_gas_mass}) and average velocity dispersion of 400 \kms\ by isolating the regions of low ionization and high-velocity dispersion ($V_{\sigma}>250$ \kms). The estimated kinetic energy is for the total turbulent region that includes the ISM and CGM of the galaxy. When estimating the total ionized gas mass over the high-velocity dispersion region, we use an electron density of 100\eden\ that is measured using the \sii\ line ratios and assuming an electron temperature of 1$\times10^{4}$ K using the PyNeb package \citep{Luridiana15}. We attempt to estimate the electron density on both ISM ($<2.6$ kpc) and larger CGM scales and find similar values within the assumed gas conditions, given our signal-to-noise ratio. Assuming a maximum radial extent for the shocked gas of 10 kpc and a shock propagation velocity similar to the velocity dispersion implies a dynamical time scale of 25 Myr, and a kinetic turbulence power of $\sim3\times10^{43}$\ergs. The driving mechanism behind the large-scale shocks in the W2246-0526 needs to inject at least this amount of kinetic power into the surrounding warm ionized gas to drive the observed shocks. We estimate that approximately 50\% of the ionized turbulent gas mass is within the ISM of the quasar host galaxy, inside 2 times the effective stellar radius of the galaxy (2.6 kpc), while the rest of the ionized turbulent gas lies on larger CGM scales.

\subsection{The nuclear outflow}
In the nuclear ($r<1.2$ kpc) region of the hot dust-obscured galaxy, we detect a powerful fast moving ionized outflow with velocities reaching $13,000$ \kms. This high velocity outflow is most likely driven by the quasar and is one of the fastest measured outflows detected in emission \citep{Jun17,perr19,Jun20,Finnerty20}.

\begin{figure}[!th]
    \centering
    \includegraphics[width=3.4in]{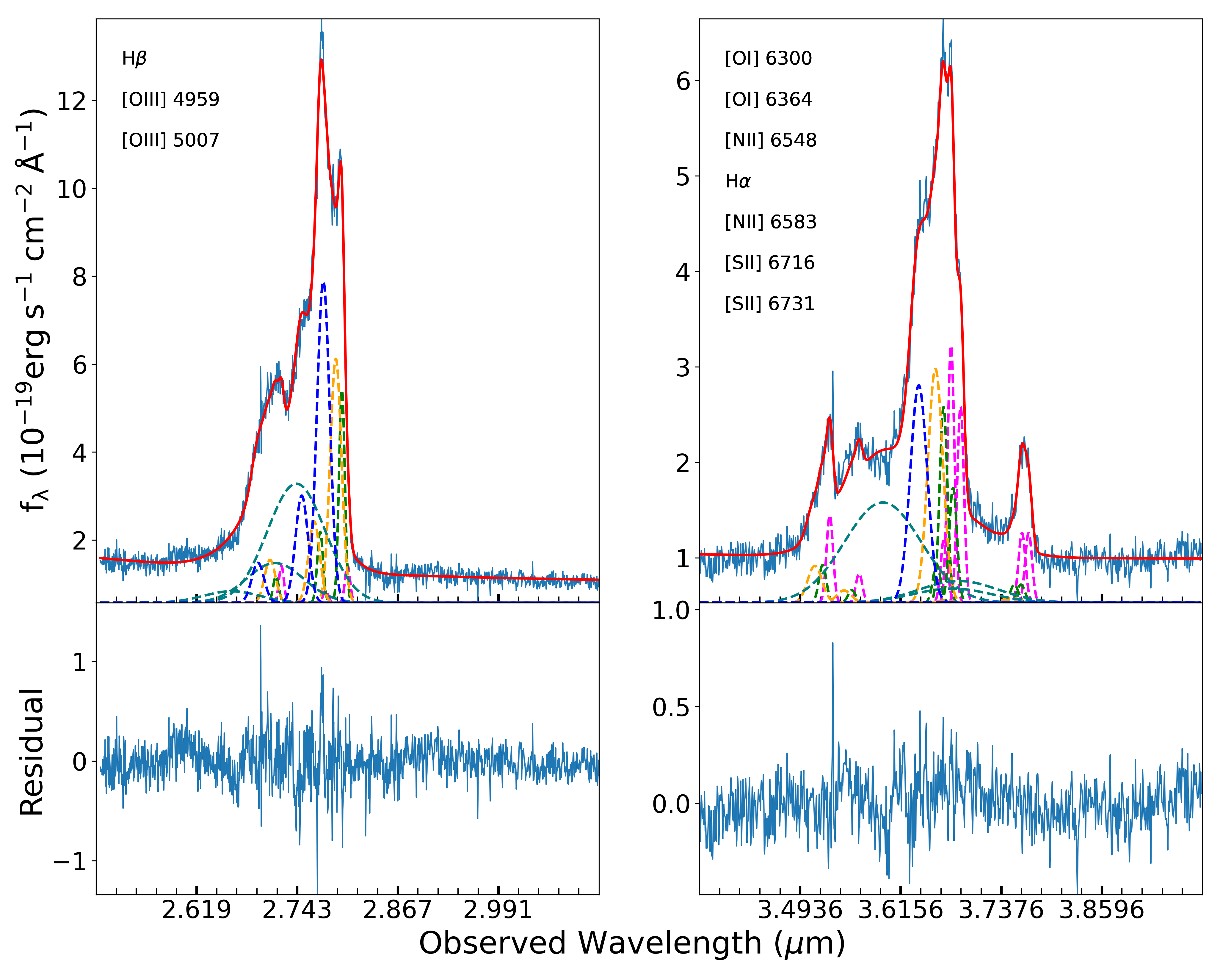}\\
    \includegraphics[width=3.4in]{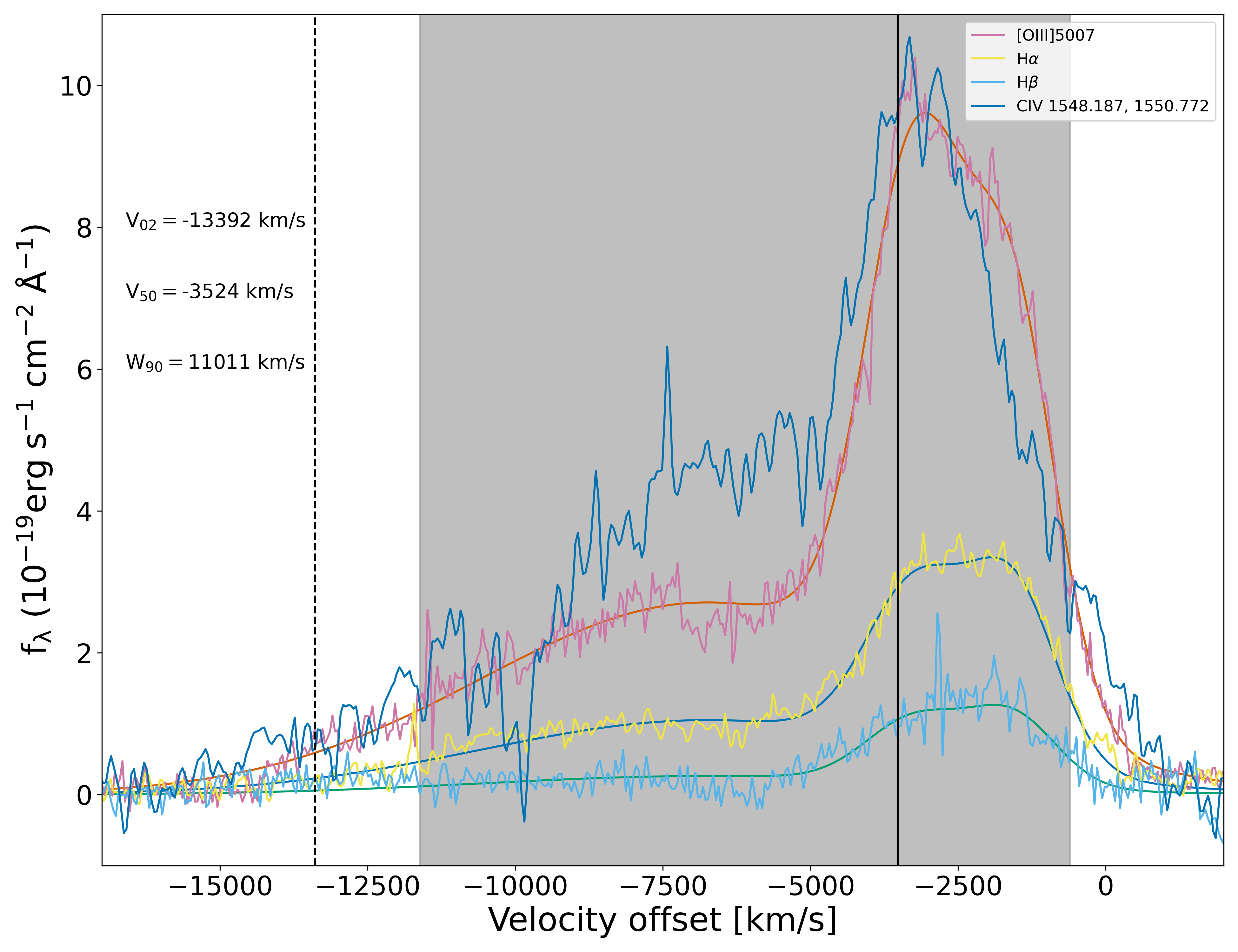}
    \caption{Top panels show the nuclear spectrum. The left panel shows the \hb\ and \oiii\ complex, while the top right panel shows the \oi, \nii, \ha, and \sii\ complex. The red curve shows the best-fit \qfit\ model. The dashed lines show the individual kinematic components for each line, where each kinematic component has a unique color. Below each complex, we show the residuals after subtracting the emission line and continuum models. The bottom panel shows the best fit \oiii, \hb\, and \ha\ line profiles of the broad blueshifted kinematic components, with the rest of the models subtracted out. We overlay the rest-frame UV \civ\ emission line profile for comparison with the optical emission lines. The velocities on the x-axis are measured relative to the systemic redshift measured using the \cii\ 158\micron\ emission line \citep{diaz18}. The gray-shaded region marks the velocity range containing 90\% of the integrated flux for the broad \oiii\ blueshifted kinematic components. The solid black line shows the average velocity of the profile (V$_{50}$), and the dashed black line shows the maximum velocity (V$_{02}$) where 2\% of the flux is integrated on the blue wing.} 
    \label{fig:nuclear-fit}
\end{figure}

To estimate the energetics, we need to measure the mass of gas in the high-velocity component. We extract a spectrum using a 2.5-pixel radius circle aperture centered on the hot dust-obscured galaxy. We apply an aperture correction per channel to the spectrum based on measurements from a bright flux calibration star that was reduced in the exact same manner as the science data (program ID 3399). We fit the \hb, \oiii, \oi, \ha, \nii\ and \sii\ lines with a combination of narrow ($<300$ \kms) and broad ($>300$ \kms) Gaussians to account for gas in gravitational motion in the galaxy and an outflow. The width and redshift of each kinematic component are tied to the \oiii\ line as it has the highest signal-to-noise ratio in the nuclear spectrum, we limit the range of allowed line ratios between \ha\ and \hb\ to satisfy case B recombination. The final best-fit spectrum consists of three broad and highly blueshifted Gaussians and two narrower components for the host galaxy near systemic velocity. We do not associate any meaning to the individual Gaussians. The number of components is selected to explain as many emission line profiles as possible with a common profile, but specifically focusing on explaining the \oiii\ and \ha\ line profiles while also satisfying case B recombination between \ha\ and \hb\ with the smallest $\chi^{2}$ value. We present the best-fit model to the nuclear spectrum in Figure \ref{fig:nuclear-fit}. In the bottom of \ref{fig:nuclear-fit} we show the profiles associated with broad blueshifted emission from \oiii, \ha, and \hb, we plot line profiles' intensity against the velocity relative to the systemic redshift of W2246-0526 measured from the \cii\ 158 \micron\ line \citep{diaz18}. In addition, we overlay the \civ\ $\lambda \lambda$ 1548 \AA, 1550\AA\ emission line spectrum from rest-frame UV observations taken by the MUSE instrument (Shobhana et al. in-prep.), extracted over the nuclear region. The \civ\ profile shows a similar shape to the rest-frame optical emission lines. Given the extremely broad ($W_{90}$=11,000 \kms) and blueshifted ($V_{50}=-3524$\kms) nature of these lines, it indicates that no realistic gravitational potential can contain the emitting ionized gas. The detection of \civ\ and \oiii\ indicates that the gas is likely photoionized by the quasar; however, a significant fraction of the \civ\ can arise from scattering as has been seen in several heavily dust-obscured sources \citep{alex18}. We attempted to fit the spectrum associating the broadest \ha\ emission with a broad-line region model, where there is no associated broad component in any of the forbidden lines. However, that fit provided an unsatisfactory fit to both the \oiii\ and \hb\ lines, with the redshift of the broad-line region having an offset of -5000-6000 \kms\ from the systemic redshift, and the line ratios between \ha\ and \hb\ significantly outside what is allowed by case B recombination. We find that the broadest emission seen in \ha\ has an accompanying component in \oiii\ but also in several other forbidden lines, indicating that this gas is beyond the broad-line region. The emitting gas producing these lines has to be on scales of a few 10s to 100s pc since the gas on smaller scales is likely to be collisionally de-excited \citep{hama11} and would not strongly emit in forbidden lines. To estimate the uncertainties, we run an MCMC algorithm with 50000 steps and discarding 10\% of the chain.

We measure a total \ha\ flux of 2.7$\times10^{-16}$\ferg\ and a luminosity of 6$\times10^{43}$ \ergs\ summing the integrated line flux of the broad blueshifted velocity kinematic components. Unfortunately, the electron density is difficult to measure for the outflow region as the \sii\ doublet is blended because the lines are broad. We assume a range of electron density of 500-1000 \eden\ for the energetics calculation. The electron density is within the range of measured values in other Hot DOGs \citep{Jun20}.

Using equation \ref{equation:ionized_gas_mass} we estimate a total ionized gas mass of $(0.2\pm0.1)\times10^{9}$ \msun. We use the following equation to estimate the outflow rate:

\begin{equation}\label{equation:outflow-simple}
    \dot{M}{_{ionized}}=\frac{M_{ionized}v_{out}}{R},
\end{equation}

\noindent where $M_{ionized}$ is the ionized gas mass, $v_{out}$ is the velocity of the outflow and $R$ is the radius of the outflow. For the outflow velocity, we use the blueshifted wing of the emission line to measure the maximum velocity since, due to projection effects, the wings of the emission lines likely show the true velocity of the outflowing material \citep{cano12,gree12}. We use the velocity where 10$\%$ of the line is integrated as the outflow velocity. When computing the velocity, we only use the combined model of the broad Gaussian components. The outflow velocity is computed by setting the following equation to 0.1:

\begin{equation}
    F(v)=\int\limits_{-\infty}^{v}f(v')/\int\limits_{-\infty}^{\infty}f(v')dv'
\end{equation}

\noindent For outflow radius, we use PSF FWHM/2, since it is spatially unresolved. We measure an outflow rate of 2000$\pm$800\myr\ and a kinetic luminosity of $(6\pm2)\times10^{46} \rm~erg~s^{-1}$ using the following equation:

\begin{equation}
    L_{\rm kinetic} = \frac{1}{2} \dot{M}_{{ionized}}\times v_{out}^{2}
\end{equation}

\noindent We present the measured properties for the nuclear outflow and their estimated uncertainties in Table \ref{tab:nuclear_outflow_properties}. In Figure \ref{fig:nuclear-fit}, we also present additional non-parametric values for the outflow, including $W_{90}$, which measures the width of the line containing 90\% of the total flux, the maximum blueshifted velocity ($V_{02}$) and the average velocity ($V_{50}$) of the emission associated with the broadest components relative to the systemic redshift of the W2246-0526.

\begin{table}[!th]
    \centering
    \begin{tabular}{c|c|c}
        Parameter    & value         & unit\\
        \hline
        \hline
                     & Shock Region &\\ 
        \hline
        logL(\ha)    &  44.05$\pm$0.04 & \ergs\\
        $<v_\sigma>$ &404 $\pm$ 40 & \kms\\
        M$^{ionized}$ & $(6\pm2)\times10^{9}$ ($ \frac{\rm 100 cm^{-3}}{n_{e}}$) & \msun \\
        E$_{turb}$    & $(3\pm1)\times10^{58}$ & erg\\
        L$_{turb}$    & $(3\pm1)\times10^{43}$ & \ergs\\
        t$_{dynamical}$ & 25 & Myr\\
        \hline
        \hline \\
                     & Nuclear Outflow &\\
        \hline 
        logL(\ha)     & 43.65$\pm$0.04 & \ergs \\
        W$_{90}$      &   11011 & \kms \\
        V$\rm_{out}$  & 9961    & \kms \\
        M$^{ionized}$ & $(2\pm1)\times10^{8}$ ($ \frac{\rm 1000 cm^{-3}}{n_{e}}$) & \msun \\
        $\dot{M}$    &  2000 $\pm$800 ($ \frac{\rm 1000 cm^{-3}}{n_{e}}$)  & \myr \\
        log10($\dot{P}_{outflow}$) &   38$\pm$0.2                                           & g cm s$^{-1}$ \\
        log10($\dot{E}_{outflow}$) &   46.8$\pm$0.2                                           & \ergs       \\
        $\frac{\dot{E}_{outflow}}{L_{AGN}}$    & 4.5$\pm$2 ($ \frac{\rm 1000 cm^{-3}}{n_{e}}$)       &                    \%       \\
        $\frac{\dot{P}_{outflow}}{\dot{P}_{AGN}}$ & 3$\pm$1 ($ \frac{\rm 1000 cm^{-3}}{n_{e}}$)      &                \\

    \end{tabular}
    \caption{Measured properties of the kinematics, dynamics, and energetics of the shocked region and the nuclear outflow in W2246-0526.}
    \label{tab:nuclear_outflow_properties}

\end{table}

\section{Discussion}
\label{sec:disc}

\subsection{The source behind the galaxy scale shocks}

We estimate a mass outflow rate of 2000 \myr\ and an outflow kinetic power of $6\times10^{46}$ \ergs, which is more than three orders of magnitude larger than the turbulent kinetic power measured in the socked circumgalactic medium. We estimate a coupling efficiency of $\sim$ 5\% between the bolometric luminosity of the quasar and the kinetic luminosity of the outflow, which is above the minimum coupling efficiency needed to provide negative quasar feedback in simulations \citep{choi15,Costa18,Mercedes-Feliz23,Wellons23}. We measure a momentum flux ratio between the ionized outflow ($1.2\times10^{38}$ g cm s$^{-1}$) and the quasar accretion disk ($4.5\times10^{37}$ g cm s$^{-1}$) of $\sim$ 3, consistent with either an outflow driven by radiation pressure on dust grains \citep{thom15,Costa18} or through an energy-conserving shock \citep{fauc12a,fauc12b}. \\

The quasar-driven outflow in the center of the system clearly has a sufficient kinetic luminosity to power the observed galaxy-scale shocks, plus the circumgalactic medium-scale shocks. To investigate whether other sources of feedback, such as supernova explosions, can drive the shocks, we need to compare the amount of energy that supernova explosions can deposit into the interstellar and circumgalactic medium. We assume each supernova explodes with an energy of $E_{SN}=10^{51}$ erg with one explosion every 100 years for a star formation rate of 1 \myr, leading to the following estimate of the energy deposition into the interstellar and circumgalactic medium:

\begin{equation}
    \dot{E}_{SNe} = \xi E_{SN} \dot{M}_{SFR}f_{SN} \sim 3\times10^{40} \frac{\dot{M}_{SFR}}{1M_{\odot}yr^{-1}} \rm erg~s^{-1} \label{eq:SNe_energy}
\end{equation}

\noindent where $\xi$ is the energy coupling between each supernova explosion and the interstellar and circumgalactic medium, taken to be 10\% \citep{Thornton98,murr05}. To drive the observed shocks in the interstellar and circumgalactic medium requires an average star formation rate of 1000 \myr\ over the past 25 Myr. Currently, within 10 kpc from the quasar, there is no strong evidence for photoionization consistent purely with star formation. We see a broad distribution of the \niiha\ and \siiha\ line ratios where the lower ratios at the lowest velocity dispersion are more consistent with star formation, and the highest line ratios with the highest velocity dispersion are more consistent with shock excitation. The overlapping region near the minima of the distribution (Figure \ref{fig:hist-ratios}) may be attributed to a transitional region between photoionization by massive stars and shocks. This range of line ratios is consistent with a skewed velocity dispersion distribution towards the lower ($<300$ \kms) velocity dispersion values in the 10 kpc radius around the quasar, hence overlapping with both lower velocity dispersion and line ratios, indicating a true transitional region between star formation and shock excitation. However, it should be noted that the line ratios are still, on average, higher than what is seen in pure star-forming galaxies; there is still likely a significant contribution from radiative shocks in these spaxels. We can place an upper limit on the recent unobscured star formation rate by integrating the \ha\ flux over the spaxels that are within $\pm$0.1 dex around the minima of the \niiha\ line ratio distribution (Figure \ref{fig:hist-ratios}) and convert it into a star formation rate using the empirical \ha\ luminosity star formation rate relation \citep{kenn98}. The 0.1 dex range selected extends to the highest \niiha\ line ratios observed in pure star-forming galaxies at $z\sim2$ \citep{Strom17}. We place an upper limit on the dust-corrected (using Balmer decrement) recent star formation rate of $<$100 \myr\ integrated over 10 kpc radius surrounding the quasar. Most of the spaxels in the transitional region between star formation and shock excitation overlap with the rest-frame UV continuum emission from the quasar host galaxy \citep{Fan18}, with additional emission extending towards the west. The rest of the spaxels consistent with photoionization by massive stars are located $\sim$15-25 kpc away from the quasar host galaxy and are unlikely to contribute to driving the shocks in the inner 10 kpc region. At most, 11\% of the total \ha\ flux is due to unobscured star formation activity within 10 kpc from the quasar, from a comparison of the total \ha\ flux integrated over the 10 kpc nebula to the \ha\ flux from spaxels that may have contribution from ionization by massive young stars.  \\

There can potentially be ongoing obscured star formation with a rate reaching 1000 \myr, since spatially integrated spectral energy distribution fitting from near to far infrared (at $\sim$50 kpc resolution) indicates the presence of cold dust that can be consistent with such a high star formation rate \citep{Sun24}. However, recent resolved ALMA observations of the far-infrared dust emission reveal that the dust in the inner 7 kpc region is at a temperature of 111$^{+24}_{-17}$ K with a total infrared luminosity of $10^{47.39}$ \ergs. This luminosity would imply an unreasonably high star formation rate of $\sim$9,000 \myr\ and, combined with the high dust temperature, implies that the dust in the inner 7 kpc is currently heated by the quasar. Beyond $\sim$ 10 kpc from the quasar, however, the ALMA data suggest there is significant ongoing obscured star formation, mostly happening in the nearby merging galaxies and tidal debris. There, the dust temperature is significantly lower, 35-40K, more similar to those seen in starburst regions in nearby dusty galaxies \citep{Fernndez-Aranda25}.\\

The quasar-driven outflow requires a coupling efficiency between the kinetic energy due to turbulent motion and the kinetic power of the quasar-driven outflow of only 0.05\%. In contrast, star formation sustained at a rate of 1000 \myr\ requires a coupling efficiency of at least 10\%, nearly two orders of magnitude higher. Therefore, while star formation can contribute to the energy injection into the interstellar and circumgalactic medium, the quasar is far more likely to be the primary driver of the shocks. Shocks in mergers are commonly observed in luminous and ultra-luminous infrared galaxies in the nearby universe \citep{rich11,Mortazavi19}. However, often, the elevated line ratios are seen at maximum velocity dispersions of about 200 \kms, whereas for W2246-0526, we observe velocity dispersions with elevated line ratios out to 700 \kms (Figure \ref{fig:ratio_vs_dispersion}) with a distribution that is more skewed to larger velocity values compared to local massive galaxy mergers \citep{rich11,Mortazavi19}. Furthermore, the tidal features that connect the nearby companion galaxies to the central galaxy do not show evidence of emission due to fast shocks; rather, this emission is concentrated in a 10 kpc nebula around the quasar.\\

In summary, while merger activity and star formation can be a source of feedback, the outflow driven by the central quasar has the most kinetic energy available to induce the observed shocks, making it the most likely source. This provides strong evidence that quasars prevent gas from efficiently cooling to form stars and accrete gas from the inner circumgalactic medium to galaxy scales. 

\subsection{Discussion of the high gas density in the CGM of W2246-0526}

The observed electron densities are quite high for gas in the CGM around W2246-0526. However, the environment around W2246-0526 contains many satellite galaxies that are dynamically interacting with the central massive galaxy. Gas stripping from the ISM of these galaxies can contribute highly dense cold clumps into the surrounding CGM. Recent hydrodynamical simulations predict that a large fraction of the gas in the CGM comes from the stripping of gas from nearby satellite galaxies \citep{Angles-Alcazar17}. It is likely that due to sensitivity effects, we are picking up the densest clouds in the CGM. Hence, we are biased towards the denser regions since the line emission from these dense clouds is proportional to the density squared. Large hydrogen column densities surrounding luminous quasar measured through absorption line studies ($log_{10}(N_{H})\sim20.5\pm1$, \citet{Lau16}) combined with large \ly\ surface brightness of up to 1$\times10^{-17}$ \surff\ in the inner CGM region within 50 kpc suggest relatively high electron densities $>1-10$ \eden \citep{Cai18,Cai19,Vayner23a} and further supporting the idea of dense clumps in the CGM of luminous quasars at high redshift, especially in highly overdense environments with lots of companion systems \citep{henn15}. Photoionization modeling of observed line ratios between \ly, \ha, and \heii\ $\lambda$ 1640\AA\ in the CGM of luminous quasars suggest high density, multi-phase gas in the CGM, with densities reaching those of gas in the ISM \citep{Cantalupo19,Langen23} on scales reaching up to several 100 kpc. \\

The observed high electron density in the gas on CGM scales implies a high thermal pressure of $\sim 10^{6}$ K cm$^{-3}$. Such thermal pressure is about a factor of 10 higher than the inner regions of the CGM in hydrodynamical simulations for galaxies of similar stellar mass to W2246-0526 \citep{Ramesh23}. It could be that we are overestimating the electron density by a factor of 10-100. If the true electron density is significantly lower, then the warm ionized gas is more likely to be in pressure equilibrium with the hot (T$\sim10^{6-7}$K) low-density medium. If the electron density is lower, the estimated ionized gas mass in the ISM/CGM can be a factor of 10-100 higher, and the amount of turbulent energy will also increase by the same amount. However, a factor of 100 increase in the ionized gas mass is unreasonable. The quasar-driven outflow still has sufficient energetics to explain the energy due to turbulence on the ISM/CGM scales, and the scenario becomes more likely that the quasar has to be responsible for driving the shocks, as the star formation rate would have to be even higher to achieve a reasonable coupling efficiency between the turbulent energy of the ISM/CGM and the feedback from supernovae. Alternatively, the density is not significantly lower, but other pressures are at play in keeping the warm ionized gas in pressure balance with the surrounding medium. For example, some simulations of the CGM that include cosmic rays show that the cosmic ray pressure can be 10 times higher \citep{Butsky18} than the thermal pressure, implying potential total pressures that are close to our observed value of $\sim 10^{6}$ K cm$^{-3}$.

\subsection{Comparison to other obscured quasars studied with {\it JWST}}
Given the recent observations with {\it JWST} of other obscured galaxy populations with luminous quasars, it is interesting to compare the shock-driven feedback in W2246-0526 with these other systems. Extremely red quasars are a population of powerful obscured active galactic nuclei \citep{Ishikawa21,Ma24} that often show some of the fastest moving outflows of any quasar sample \citep{perr19}. They often overlap in properties with Hot DOGs, such as rest-frame UV to infrared colors and high intrinsic infrared luminosities, and are likely part of a transitional phase in the evolution of galaxies and their supermassive black holes. Reddened quasars also often show evidence for fast-moving outflows, however not as extreme as the extremely red quasar and Hot DOG population \citep{Temple19}. As part of an early-release science program, {\it JWST} observed the extremely red quasar, J1652 ($z=2.94$) \citep{Vayner23b,Vayner24}. Similar to W2246-0526, the host galaxy of this quasar shows shocks induced by powerful quasar-driven outflows near the outskirts of the galaxy, perpendicular to the direction of the outflow. The W2246-0526 and J1652 have similar numbers of merging satellite galaxies, similar stellar mass, both show strong nuclear obscuration, and both have powerful outflows, a common property of both extremely red quasars and hot dust-obscured galaxies \citep{perr19,Jun20}. The \ha\ luminosity due to shocks in the W2246-0526 system of $1\times10^{44}$ \ergs is about a factor of 50 higher than the J1652 system. The bolometric luminosity of the W2246-0526 quasar is only about a factor 2.5 higher, while the outflow kinetic luminosity and momentum flux are about a factor of 10 higher. Larger energetics in the W2246-0526 may explain the higher \ha\ luminosity of the shocked gas in the W2246-0526 system. Higher nuclear obscuration can also increase the detected \ha\ luminosity due to shocks. In fact, in the J1652 system, shocks are only detected in regions outside the quasar photoionization cone; likely, shocks are present in the region photoionized by the quasar but are heavily diluted by the radiation from quasar ionization. In the W2246-0526 system, we see nearly isotropic shock-dominated emission out to 10 kpc from the central quasar, which would suggest that the dust in the host galaxy is likely distributed homogeneously with a large covering factor around the quasar. This could explain why the gas on galactic and circumgalactic scales does not show signatures of being photoionized by the quasar itself but instead displays low ionization emission line ratios typical of radiative shocks. Indeed, multi-wavelength observations of the W2246-0526 quasar indicate that the covering factor of the dust-obscuring material in the nuclear region is close to unity \citep{Fan16b}. \\

\section{Conclusions}
\label{sec:conc}

In this paper, we present {\it JWST} NIRSpec mosaic IFU observations of W2246-0526, a dust-obscured galaxy hosting a luminous quasar at $z=4.601$. W2246-0526 is the most luminous obscured quasar known to date. We mapped rest-frame optical emission lines (\oii, \hb, \oiii, \oi, \ha, \nii\ and \sii) at kpc-scale resolution out to 60 kpc from the quasar, and we find: \\

$\bullet$ Filamentary structures in ionized emission that connect the nearby companion galaxies to the host galaxy of the central luminous quasar. \\

$\bullet$ In the central 10 kpc surrounding the quasar, an extended nebula of low-ionization gas that we find to be heated by radiative shocks. Shocks are detected out to eight times the effective radius of the stellar light of the quasar host galaxy, indicating for the first time the presence of mechanical heating of gas on circumgalactic scales at this cosmic epoch. \\

$\bullet$ A powerful quasar-driven outflow in the inner 1.2 kpc from the quasar with velocities reaching 13,000 \kms, the fastest outflow detected in emission to date. We measure an outflow rate of 2000 \myr\ and a kinetic luminosity of 6$\times10^{46}$ \ergs with a 5\% coupling efficiency to the bolometric luminosity of the quasar. The outflow is powerful enough to explain the kinetic energy due to the turbulent motion of the gas observed on galactic and circumgalactic scales out to 10 kpc, suggesting it is the primary driver of the observed shocks. \\

$\bullet$  While star formation activity can partially contribute to driving the radiative shocks, the star formation rate required to be the dominant heating mechanism is much larger than what is measured within the low-ionization nebula around 10 kpc from the quasar. Thus, the most likely scenario is that the quasar provides most of the energy and momentum necessary to drive the shocks, with likely only a minor contribution from star formation. \\

In summary, the powerful quasar in W2246-0526 is likely driving the turbulence and shocks observed on large scales around its host, probably preventing gas from cooling efficiently and being accreted to the central kpc of the system. If W2246-0526 is representative of the Hot DOG population, these results suggest that luminous, obscured quasars can provide effective negative feedback not only on galactic but also on CGM scales as early as 1.3 Gyr after the Big Bang. \\

\begin{acknowledgments}
This work is based on observations made with the NASA/ESA/CSA James Webb Space Telescope. The data were obtained from the Mikulski Archive for Space Telescopes at the Space Telescope Science Institute, which is operated by the Association of Universities for Research in Astronomy, Inc., under NASA contract NAS 5-03127 for JWST. These observations are associated with program 1712.\\

\noindent Portions of this research were carried out at the Jet Propulsion Laboratory, California Institute of Technology, under a contract with the National Aeronautics and Space Administration.
RJA was supported by FONDECYT grant number 1231718 and by the ANID BASAL project FB210003.\\

\noindent DAA acknowledges support by NSF grant AST-2108944, NASA grant ATP23-0156, STScI grants JWST-GO-01712.009-A and JWST-AR-04357.001-A, Simons Foundation Award CCA-1018464, and Cottrell Scholar Award CS-CSA-2023-028 by the Research Corporation for Science Advancement.\\

\noindent TDS acknowledges the research project was supported by the Hellenic Foundation for Research and Innovation (HFRI) under the ``2nd Call for HFRI Research Projects to support Faculty Members \& Researchers" (Project Number: 3382).\\

\noindent The authors would like to thank the anonymous referee for their excellent comments that helped improve the manuscript. 

\end{acknowledgments}

\vspace{5mm}
\facilities{JWST(NIRSpec), HST(WFC3) }\\
The data is available at MAST: \dataset[10.17909/zydy-1534]{\doi{10.17909/zydy-1534}}
\software{astropy \citep{Astropy2013, Astropy2018},  reproject \citep{thomas_robitaille_2023_7584411}, \qfit \citep{ifsfit2014}}

\bibliography{bib.bib}{}

\begin{thebibliography}{}
\expandafter\ifx\csname natexlab\endcsname\relax\def\natexlab#1{#1}\fi
\providecommand{\url}[1]{\href{#1}{#1}}
\providecommand{\dodoi}[1]{doi:~\href{http://doi.org/#1}{\nolinkurl{#1}}}
\providecommand{\doeprint}[1]{\href{http://ascl.net/#1}{\nolinkurl{http://ascl.net/#1}}}
\providecommand{\doarXiv}[1]{\href{https://arxiv.org/abs/#1}{\nolinkurl{https://arxiv.org/abs/#1}}}

\bibitem[{{Alarie} \& {Morisset}(2019)}]{Alarie19}
{Alarie}, A., \& {Morisset}, C. 2019, \rmxaa, 55, 377, \dodoi{10.22201/ia.01851101p.2019.55.02.21}

\bibitem[{{Alexandroff} {et~al.}(2018){Alexandroff}, {Zakamska}, {Barth}, {Hamann}, {Strauss}, {Krolik}, {Greene}, {P{\^a}ris}, \& {Ross}}]{alex18}
{Alexandroff}, R.~M., {Zakamska}, N.~L., {Barth}, A.~J., {et~al.} 2018, \mnras, 479, 4936, \dodoi{10.1093/mnras/sty1685}

\bibitem[{{Allen} {et~al.}(2008){Allen}, {Groves}, {Dopita}, {Sutherland}, \& {Kewley}}]{alle08}
{Allen}, M.~G., {Groves}, B.~A., {Dopita}, M.~A., {Sutherland}, R.~S., \& {Kewley}, L.~J. 2008, \apjs, 178, 20, \dodoi{10.1086/589652}

\bibitem[{{Anderson} {et~al.}(2016){Anderson}, {Churazov}, \& {Bregman}}]{Anderson16}
{Anderson}, M.~E., {Churazov}, E., \& {Bregman}, J.~N. 2016, \mnras, 455, 227, \dodoi{10.1093/mnras/stv2314}

\bibitem[{{Angl{\'e}s-Alc{\'a}zar} {et~al.}(2017){Angl{\'e}s-Alc{\'a}zar}, {Faucher-Gigu{\`e}re}, {Kere{\v{s}}}, {Hopkins}, {Quataert}, \& {Murray}}]{Angles-Alcazar17}
{Angl{\'e}s-Alc{\'a}zar}, D., {Faucher-Gigu{\`e}re}, C.-A., {Kere{\v{s}}}, D., {et~al.} 2017, \mnras, 470, 4698, \dodoi{10.1093/mnras/stx1517}

\bibitem[{{Angl{\'e}s-Alc{\'a}zar} {et~al.}(2021){Angl{\'e}s-Alc{\'a}zar}, {Quataert}, {Hopkins}, {Somerville}, {Hayward}, {Faucher-Gigu{\`e}re}, {Bryan}, {Kere{\v{s}}}, {Hernquist}, \& {Stone}}]{Angles-Alcazar21}
{Angl{\'e}s-Alc{\'a}zar}, D., {Quataert}, E., {Hopkins}, P.~F., {et~al.} 2021, \apj, 917, 53, \dodoi{10.3847/1538-4357/ac09e8}

\bibitem[{{Arrigoni Battaia} {et~al.}(2019){Arrigoni Battaia}, {Hennawi}, {Prochaska}, {O{\~n}orbe}, {Farina}, {Cantalupo}, \& {Lusso}}]{Arrigoni-Battaia19}
{Arrigoni Battaia}, F., {Hennawi}, J.~F., {Prochaska}, J.~X., {et~al.} 2019, \mnras, 482, 3162, \dodoi{10.1093/mnras/sty2827}

\bibitem[{{Astropy Collaboration} {et~al.}(2013){Astropy Collaboration}, {Robitaille}, {Tollerud}, {Greenfield}, {Droettboom}, {Bray}, {Aldcroft}, {Davis}, {Ginsburg}, {Price-Whelan}, {Kerzendorf}, {Conley}, {Crighton}, {Barbary}, {Muna}, {Ferguson}, {Grollier}, {Parikh}, {Nair}, {Unther}, {Deil}, {Woillez}, {Conseil}, {Kramer}, {Turner}, {Singer}, {Fox}, {Weaver}, {Zabalza}, {Edwards}, {Azalee Bostroem}, {Burke}, {Casey}, {Crawford}, {Dencheva}, {Ely}, {Jenness}, {Labrie}, {Lim}, {Pierfederici}, {Pontzen}, {Ptak}, {Refsdal}, {Servillat}, \& {Streicher}}]{Astropy2013}
{Astropy Collaboration}, {Robitaille}, T.~P., {Tollerud}, E.~J., {et~al.} 2013, \aap, 558, A33, \dodoi{10.1051/0004-6361/201322068}

\bibitem[{{Astropy Collaboration} {et~al.}(2018){Astropy Collaboration}, {Price-Whelan}, {Sip{\H{o}}cz}, {G{\"u}nther}, {Lim}, {Crawford}, {Conseil}, {Shupe}, {Craig}, {Dencheva}, {Ginsburg}, {VanderPlas}, {Bradley}, {P{\'e}rez-Su{\'a}rez}, {de Val-Borro}, {Aldcroft}, {Cruz}, {Robitaille}, {Tollerud}, {Ardelean}, {Babej}, {Bach}, {Bachetti}, {Bakanov}, {Bamford}, {Barentsen}, {Barmby}, {Baumbach}, {Berry}, {Biscani}, {Boquien}, {Bostroem}, {Bouma}, {Brammer}, {Bray}, {Breytenbach}, {Buddelmeijer}, {Burke}, {Calderone}, {Cano Rodr{\'\i}guez}, {Cara}, {Cardoso}, {Cheedella}, {Copin}, {Corrales}, {Crichton}, {D'Avella}, {Deil}, {Depagne}, {Dietrich}, {Donath}, {Droettboom}, {Earl}, {Erben}, {Fabbro}, {Ferreira}, {Finethy}, {Fox}, {Garrison}, {Gibbons}, {Goldstein}, {Gommers}, {Greco}, {Greenfield}, {Groener}, {Grollier}, {Hagen}, {Hirst}, {Homeier}, {Horton}, {Hosseinzadeh}, {Hu}, {Hunkeler}, {Ivezi{\'c}}, {Jain}, {Jenness}, {Kanarek}, {Kendrew}, {Kern}, {Kerzendorf}, {Khvalko}, {King}, {Kirkby}, {Kulkarni},
  {Kumar}, {Lee}, {Lenz}, {Littlefair}, {Ma}, {Macleod}, {Mastropietro}, {McCully}, {Montagnac}, {Morris}, {Mueller}, {Mumford}, {Muna}, {Murphy}, {Nelson}, {Nguyen}, {Ninan}, {N{\"o}the}, {Ogaz}, {Oh}, {Parejko}, {Parley}, {Pascual}, {Patil}, {Patil}, {Plunkett}, {Prochaska}, {Rastogi}, {Reddy Janga}, {Sabater}, {Sakurikar}, {Seifert}, {Sherbert}, {Sherwood-Taylor}, {Shih}, {Sick}, {Silbiger}, {Singanamalla}, {Singer}, {Sladen}, {Sooley}, {Sornarajah}, {Streicher}, {Teuben}, {Thomas}, {Tremblay}, {Turner}, {Terr{\'o}n}, {van Kerkwijk}, {de la Vega}, {Watkins}, {Weaver}, {Whitmore}, {Woillez}, {Zabalza}, \& {Astropy Contributors}}]{Astropy2018}
{Astropy Collaboration}, {Price-Whelan}, A.~M., {Sip{\H{o}}cz}, B.~M., {et~al.} 2018, \aj, 156, 123, \dodoi{10.3847/1538-3881/aabc4f}

\bibitem[{{Behroozi} {et~al.}(2010){Behroozi}, {Conroy}, \& {Wechsler}}]{Behroozi10}
{Behroozi}, P.~S., {Conroy}, C., \& {Wechsler}, R.~H. 2010, \apj, 717, 379, \dodoi{10.1088/0004-637X/717/1/379}

\bibitem[{{B{\"o}ker} {et~al.}(2022){B{\"o}ker}, {Arribas}, {L{\"u}tzgendorf}, {Alves de Oliveira}, {Beck}, {Birkmann}, {Bunker}, {Charlot}, {de Marchi}, {Ferruit}, {Giardino}, {Jakobsen}, {Kumari}, {L{\'o}pez-Caniego}, {Maiolino}, {Manjavacas}, {Marston}, {Moseley}, {Muzerolle}, {Ogle}, {Pirzkal}, {Rauscher}, {Rawle}, {Rix}, {Sabbi}, {Sargent}, {Sirianni}, {te Plate}, {Valenti}, {Willott}, \& {Zeidler}}]{Boker22}
{B{\"o}ker}, T., {Arribas}, S., {L{\"u}tzgendorf}, N., {et~al.} 2022, \aap, 661, A82, \dodoi{10.1051/0004-6361/202142589}

\bibitem[{{Borisova} {et~al.}(2016){Borisova}, {Cantalupo}, {Lilly}, {Marino}, {Gallego}, {Bacon}, {Blaizot}, {Bouch{\'e}}, {Brinchmann}, {Carollo}, {Caruana}, {Finley}, {Herenz}, {Richard}, {Schaye}, {Straka}, {Turner}, {Urrutia}, {Verhamme}, \& {Wisotzki}}]{bori16}
{Borisova}, E., {Cantalupo}, S., {Lilly}, S.~J., {et~al.} 2016, \apj, 831, 39, \dodoi{10.3847/0004-637X/831/1/39}

\bibitem[{Bradley {et~al.}(2023)Bradley, Sip{\H o}cz, Robitaille, Tollerud, Vin{\'{\i}}cius, Deil, Barbary, Wilson, Busko, Donath, G{\"u}nther, Cara, Lim, Me{\ss}linger, Conseil, Bostroem, Droettboom, Bray, Bratholm, Barentsen, Craig, Rathi, Pascual, Perren, Georgiev, de~Val-Borro, Kerzendorf, Bach, Quint, \& Souchereau}]{larry_bradley_2023_7946442}
Bradley, L., Sip{\H o}cz, B., Robitaille, T., {et~al.} 2023, astropy/photutils: 1.8.0, 1.8.0,  Zenodo, \dodoi{10.5281/zenodo.7946442}

\bibitem[{{Butsky} \& {Quinn}(2018)}]{Butsky18}
{Butsky}, I.~S., \& {Quinn}, T.~R. 2018, \apj, 868, 108, \dodoi{10.3847/1538-4357/aaeac2}

\bibitem[{{Cai} {et~al.}(2018){Cai}, {Hamden}, {Matuszewski}, {Prochaska}, {Li}, {Cantalupo}, {Arrigoni Battaia}, {Martin}, {Neill}, {O'Sullivan}, {Wang}, {Moore}, \& {Morrissey}}]{Cai18}
{Cai}, Z., {Hamden}, E., {Matuszewski}, M., {et~al.} 2018, \apjl, 861, L3, \dodoi{10.3847/2041-8213/aacce6}

\bibitem[{{Cai} {et~al.}(2019){Cai}, {Cantalupo}, {Prochaska}, {Arrigoni Battaia}, {Burchett}, {Li}, {Chisholm}, {Bundy}, \& {Hennawi}}]{Cai19}
{Cai}, Z., {Cantalupo}, S., {Prochaska}, J.~X., {et~al.} 2019, \apjs, 245, 23, \dodoi{10.3847/1538-4365/ab4796}

\bibitem[{{Calzetti} {et~al.}(2000){Calzetti}, {Armus}, {Bohlin}, {Kinney}, {Koornneef}, \& {Storchi-Bergmann}}]{Calzetti2000}
{Calzetti}, D., {Armus}, L., {Bohlin}, R.~C., {et~al.} 2000, \apj, 533, 682, \dodoi{10.1086/308692}

\bibitem[{{Cano-D{\'{\i}}az} {et~al.}(2012){Cano-D{\'{\i}}az}, {Maiolino}, {Marconi}, {Netzer}, {Shemmer}, \& {Cresci}}]{cano12}
{Cano-D{\'{\i}}az}, M., {Maiolino}, R., {Marconi}, A., {et~al.} 2012, \aap, 537, L8, \dodoi{10.1051/0004-6361/201118358}

\bibitem[{{Cantalupo} {et~al.}(2005){Cantalupo}, {Porciani}, {Lilly}, \& {Miniati}}]{cant05}
{Cantalupo}, S., {Porciani}, C., {Lilly}, S.~J., \& {Miniati}, F. 2005, \apj, 628, 61, \dodoi{10.1086/430758}

\bibitem[{{Cantalupo} {et~al.}(2019){Cantalupo}, {Pezzulli}, {Lilly}, {Marino}, {Gallego}, {Schaye}, {Bacon}, {Feltre}, {Kollatschny}, {Nanayakkara}, {Richard}, {Wendt}, {Wisotzki}, \& {Prochaska}}]{Cantalupo19}
{Cantalupo}, S., {Pezzulli}, G., {Lilly}, S.~J., {et~al.} 2019, \mnras, 483, 5188, \dodoi{10.1093/mnras/sty3481}

\bibitem[{{Chen} {et~al.}(2025){Chen}, {Chen}, {Rauch}, {Vayner}, {Liu}, {Rupke}, {Greene}, {Zakamska}, {Wylezalek}, {Liu}, {Veilleux}, {Nesvadba}, \& {Bertemes}}]{Chen25}
{Chen}, M.~C., {Chen}, H.-W., {Rauch}, M., {et~al.} 2025, \apjl, 978, L18, \dodoi{10.3847/2041-8213/ad9bac}

\bibitem[{{Choi} {et~al.}(2015){Choi}, {Ostriker}, {Naab}, {Oser}, \& {Moster}}]{choi15}
{Choi}, E., {Ostriker}, J.~P., {Naab}, T., {Oser}, L., \& {Moster}, B.~P. 2015, \mnras, 449, 4105, \dodoi{10.1093/mnras/stv575}

\bibitem[{{Costa} {et~al.}(2018){Costa}, {Rosdahl}, {Sijacki}, \& {Haehnelt}}]{Costa18}
{Costa}, T., {Rosdahl}, J., {Sijacki}, D., \& {Haehnelt}, M.~G. 2018, \mnras, 479, 2079, \dodoi{10.1093/mnras/sty1514}

\bibitem[{{D'Eugenio} {et~al.}(2023){D'Eugenio}, {Perez-Gonzalez}, {Maiolino}, {Scholtz}, {Perna}, {Circosta}, {Uebler}, {Arribas}, {Boeker}, {Bunker}, {Carniani}, {Charlot}, {Chevallard}, {Cresci}, {Curtis-Lake}, {Jones}, {Kumari}, {Lamperti}, {Looser}, {Parlanti}, {Rix}, {Robertson}, {Rodriguez Del Pino}, {Tacchella}, {Venturi}, \& {Willott}}]{DEugenio23}
{D'Eugenio}, F., {Perez-Gonzalez}, P., {Maiolino}, R., {et~al.} 2023, arXiv e-prints, arXiv:2308.06317, \dodoi{10.48550/arXiv.2308.06317}

\bibitem[{{D{\'\i}az-Santos} {et~al.}(2018){D{\'\i}az-Santos}, {Assef}, {Blain}, {Aravena}, {Stern}, {Tsai}, {Eisenhardt}, {Wu}, {Jun}, {Dibert}, {Inami}, {Lansbury}, \& {Leclercq}}]{diaz18}
{D{\'\i}az-Santos}, T., {Assef}, R.~J., {Blain}, A.~W., {et~al.} 2018, Science, 362, 1034, \dodoi{10.1126/science.aap7605}

\bibitem[{{Ding} {et~al.}(2022){Ding}, {Silverman}, \& {Onoue}}]{Ding22}
{Ding}, X., {Silverman}, J.~D., \& {Onoue}, M. 2022, \apjl, 939, L28, \dodoi{10.3847/2041-8213/ac9c02}

\bibitem[{{Drake} {et~al.}(2019){Drake}, {Farina}, {Neeleman}, {Walter}, {Venemans}, {Banados}, {Mazzucchelli}, \& {Decarli}}]{Drake19}
{Drake}, A.~B., {Farina}, E.~P., {Neeleman}, M., {et~al.} 2019, \apj, 881, 131, \dodoi{10.3847/1538-4357/ab2984}

\bibitem[{{Eisenhardt} {et~al.}(2012){Eisenhardt}, {Wu}, {Tsai}, {Assef}, {Benford}, {Blain}, {Bridge}, {Condon}, {Cushing}, {Cutri}, {Evans}, {Gelino}, {Griffith}, {Grillmair}, {Jarrett}, {Lonsdale}, {Masci}, {Mason}, {Petty}, {Sayers}, {Stanford}, {Stern}, {Wright}, \& {Yan}}]{Eisenhardt12}
{Eisenhardt}, P. R.~M., {Wu}, J., {Tsai}, C.-W., {et~al.} 2012, \apj, 755, 173, \dodoi{10.1088/0004-637X/755/2/173}

\bibitem[{{Emonts} {et~al.}(2018){Emonts}, {Lehnert}, {Dannerbauer}, {De Breuck}, {Villar-Mart{\'\i}n}, {Miley}, {Allison}, {Gullberg}, {Hatch}, {Guillard}, {Mao}, \& {Norris}}]{Emonts18}
{Emonts}, B.~H.~C., {Lehnert}, M.~D., {Dannerbauer}, H., {et~al.} 2018, \mnras, 477, L60, \dodoi{10.1093/mnrasl/sly034}

\bibitem[{{Epinat} {et~al.}(2018){Epinat}, {Contini}, {Finley}, {Boogaard}, {Gu{\'e}rou}, {Brinchmann}, {Carton}, {Michel-Dansac}, {Bacon}, {Cantalupo}, {Carollo}, {Hamer}, {Kollatschny}, {Krajnovi{\'c}}, {Marino}, {Richard}, {Soucail}, {Weilbacher}, \& {Wisotzki}}]{Epinat18}
{Epinat}, B., {Contini}, T., {Finley}, H., {et~al.} 2018, \aap, 609, A40, \dodoi{10.1051/0004-6361/201731877}

\bibitem[{{Fan} {et~al.}(2018){Fan}, {Gao}, {Knudsen}, \& {Shu}}]{Fan18}
{Fan}, L., {Gao}, Y., {Knudsen}, K.~K., \& {Shu}, X. 2018, \apj, 854, 157, \dodoi{10.3847/1538-4357/aaaaae}

\bibitem[{{Fan} {et~al.}(2016){Fan}, {Han}, {Nikutta}, {Drouart}, \& {Knudsen}}]{Fan16b}
{Fan}, L., {Han}, Y., {Nikutta}, R., {Drouart}, G., \& {Knudsen}, K.~K. 2016, \apj, 823, 107, \dodoi{10.3847/0004-637X/823/2/107}

\bibitem[{{Faucher-Gigu{\`e}re} \& {Quataert}(2012)}]{fauc12b}
{Faucher-Gigu{\`e}re}, C.-A., \& {Quataert}, E. 2012, \mnras, 425, 605, \dodoi{10.1111/j.1365-2966.2012.21512.x}

\bibitem[{{Faucher-Gigu{\`e}re} {et~al.}(2012){Faucher-Gigu{\`e}re}, {Quataert}, \& {Murray}}]{fauc12a}
{Faucher-Gigu{\`e}re}, C.-A., {Quataert}, E., \& {Murray}, N. 2012, \mnras, 420, 1347, \dodoi{10.1111/j.1365-2966.2011.20120.x}

\bibitem[{{Fern{\'a}ndez Aranda} {et~al.}(2024){Fern{\'a}ndez Aranda}, {D{\'\i}az Santos}, {Hatziminaoglou}, {Assef}, {Aravena}, {Eisenhardt}, {Ferkinhoff}, {Pensabene}, {Nikola}, {Andreani}, {Vishwas}, {Stacey}, {Decarli}, {Blain}, {Brisbin}, {Charmandaris}, {Jun}, {Li}, {Liao}, {Martin}, {Stern}, {Tsai}, {Wu}, \& {Zewdie}}]{Fernandez-Aranda24}
{Fern{\'a}ndez Aranda}, R., {D{\'\i}az Santos}, T., {Hatziminaoglou}, E., {et~al.} 2024, \aap, 682, A166, \dodoi{10.1051/0004-6361/202347869}

\bibitem[{{Fern{\'a}ndez Aranda} {et~al.}(2025){Fern{\'a}ndez Aranda}, {D{\'\i}az Santos}, {Hatziminaoglou}, {Aravena}, {Stern}, {Armus}, {Assef}, {Blain}, {Charmandaris}, {Decarli}, {Eisenhardt}, {Ferkinhoff}, {Gonz{\'a}lez-L{\'o}pez}, {Jun}, {Li}, {Liao}, {Shevill}, {Shobhana}, {Tsai}, {Vayner}, {Wu}, \& {Zewdie}}]{Fernndez-Aranda25}
---. 2025, \aap, 695, L15, \dodoi{10.1051/0004-6361/202453214}

\bibitem[{{Finkelstein} {et~al.}(2023){Finkelstein}, {Bagley}, {Ferguson}, {Wilkins}, {Kartaltepe}, {Papovich}, {Yung}, {Arrabal Haro}, {Behroozi}, {Dickinson}, {Kocevski}, {Koekemoer}, {Larson}, {Le Bail}, {Morales}, {P{\'e}rez-Gonz{\'a}lez}, {Burgarella}, {Dav{\'e}}, {Hirschmann}, {Somerville}, {Wuyts}, {Bromm}, {Casey}, {Fontana}, {Fujimoto}, {Gardner}, {Giavalisco}, {Grazian}, {Grogin}, {Hathi}, {Hutchison}, {Jha}, {Jogee}, {Kewley}, {Kirkpatrick}, {Long}, {Lotz}, {Pentericci}, {Pierel}, {Pirzkal}, {Ravindranath}, {Ryan}, {Trump}, {Yang}, {Bhatawdekar}, {Bisigello}, {Buat}, {Calabr{\`o}}, {Castellano}, {Cleri}, {Cooper}, {Croton}, {Daddi}, {Dekel}, {Elbaz}, {Franco}, {Gawiser}, {Holwerda}, {Huertas-Company}, {Jaskot}, {Leung}, {Lucas}, {Mobasher}, {Pandya}, {Tacchella}, {Weiner}, \& {Zavala}}]{Finkelstein23}
{Finkelstein}, S.~L., {Bagley}, M.~B., {Ferguson}, H.~C., {et~al.} 2023, \apjl, 946, L13, \dodoi{10.3847/2041-8213/acade4}

\bibitem[{{Finnerty} {et~al.}(2020){Finnerty}, {Larson}, {Soifer}, {Armus}, {Matthews}, {Jun}, {Moon}, {Melbourne}, {Gomez}, {Tsai}, {D{\'\i}az-Santos}, {Eisenhardt}, \& {Cushing}}]{Finnerty20}
{Finnerty}, L., {Larson}, K., {Soifer}, B.~T., {et~al.} 2020, \apj, 905, 16, \dodoi{10.3847/1538-4357/abc3bf}

\bibitem[{{Galavis} {et~al.}(1997){Galavis}, {Mendoza}, \& {Zeippen}}]{Galavis97}
{Galavis}, M.~E., {Mendoza}, C., \& {Zeippen}, C.~J. 1997, \aaps, 123, 159, \dodoi{10.1051/aas:1997344}

\bibitem[{{Greene} {et~al.}(2012){Greene}, {Zakamska}, \& {Smith}}]{gree12}
{Greene}, J.~E., {Zakamska}, N.~L., \& {Smith}, P.~S. 2012, \apj, 746, 86, \dodoi{10.1088/0004-637X/746/1/86}

\bibitem[{{Groves} {et~al.}(2004){Groves}, {Dopita}, \& {Sutherland}}]{grov04a}
{Groves}, B.~A., {Dopita}, M.~A., \& {Sutherland}, R.~S. 2004, \apjs, 153, 9, \dodoi{10.1086/421113}

\bibitem[{{Gutkin} {et~al.}(2016){Gutkin}, {Charlot}, \& {Bruzual}}]{Gutkin16}
{Gutkin}, J., {Charlot}, S., \& {Bruzual}, G. 2016, \mnras, 462, 1757, \dodoi{10.1093/mnras/stw1716}

\bibitem[{{Hamann} {et~al.}(2011){Hamann}, {Kanekar}, {Prochaska}, {Murphy}, {Ellison}, {Malec}, {Milutinovic}, \& {Ubachs}}]{hama11}
{Hamann}, F., {Kanekar}, N., {Prochaska}, J.~X., {et~al.} 2011, \mnras, 410, 1957, \dodoi{10.1111/j.1365-2966.2010.17575.x}

\bibitem[{{Hennawi} {et~al.}(2015){Hennawi}, {Prochaska}, {Cantalupo}, \& {Arrigoni-Battaia}}]{henn15}
{Hennawi}, J.~F., {Prochaska}, J.~X., {Cantalupo}, S., \& {Arrigoni-Battaia}, F. 2015, Science, 348, 779, \dodoi{10.1126/science.aaa5397}

\bibitem[{{Hopkins} {et~al.}(2024){Hopkins}, {Grudic}, {Su}, {Wellons}, {Angles-Alcazar}, {Steinwandel}, {Guszejnov}, {Murray}, {Faucher-Giguere}, {Quataert}, \& {Keres}}]{Hopkins24}
{Hopkins}, P.~F., {Grudic}, M.~Y., {Su}, K.-Y., {et~al.} 2024, The Open Journal of Astrophysics, 7, 18, \dodoi{10.21105/astro.2309.13115}

\bibitem[{{Ishikawa} {et~al.}(2021){Ishikawa}, {Goulding}, {Zakamska}, {Hamann}, {Vayner}, {Veilleux}, \& {Wylezalek}}]{Ishikawa21}
{Ishikawa}, Y., {Goulding}, A.~D., {Zakamska}, N.~L., {et~al.} 2021, \mnras, 502, 3769, \dodoi{10.1093/mnras/stab137}

\bibitem[{{Jakobsen} {et~al.}(2022){Jakobsen}, {Ferruit}, {Alves de Oliveira}, {Arribas}, {Bagnasco}, {Barho}, {Beck}, {Birkmann}, {B{\"o}ker}, {Bunker}, {Charlot}, {de Jong}, {de Marchi}, {Ehrenwinkler}, {Falcolini}, {Fels}, {Franx}, {Franz}, {Funke}, {Giardino}, {Gnata}, {Holota}, {Honnen}, {Jensen}, {Jentsch}, {Johnson}, {Jollet}, {Karl}, {Kling}, {K{\"o}hler}, {Kolm}, {Kumari}, {Lander}, {Lemke}, {L{\'o}pez-Caniego}, {L{\"u}tzgendorf}, {Maiolino}, {Manjavacas}, {Marston}, {Maschmann}, {Maurer}, {Messerschmidt}, {Moseley}, {Mosner}, {Mott}, {Muzerolle}, {Pirzkal}, {Pittet}, {Plitzke}, {Posselt}, {Rapp}, {Rauscher}, {Rawle}, {Rix}, {R{\"o}del}, {Rumler}, {Sabbi}, {Salvignol}, {Schmid}, {Sirianni}, {Smith}, {Strada}, {te Plate}, {Valenti}, {Wettemann}, {Wiehe}, {Wiesmayer}, {Willott}, {Wright}, {Zeidler}, \& {Zincke}}]{Jakobsen22}
{Jakobsen}, P., {Ferruit}, P., {Alves de Oliveira}, C., {et~al.} 2022, \aap, 661, A80, \dodoi{10.1051/0004-6361/202142663}

\bibitem[{{Johnson} {et~al.}(2018){Johnson}, {Chen}, {Straka}, {Schaye}, {Cantalupo}, {Wendt}, {Muzahid}, {Bouch{\'e}}, {Herenz}, {Kollatschny}, {Mulchaey}, {Marino}, {Maseda}, \& {Wisotzki}}]{Johnson18}
{Johnson}, S.~D., {Chen}, H.-W., {Straka}, L.~A., {et~al.} 2018, \apjl, 869, L1, \dodoi{10.3847/2041-8213/aaf1cf}

\bibitem[{{Johnson} {et~al.}(2022){Johnson}, {Schaye}, {Walth}, {Li}, {Rudie}, {Chen}, {Chen}, {Epinat}, {Gaspari}, {Cantalupo}, {Kollatschny}, {Liu}, \& {Muzahid}}]{Johnson22}
{Johnson}, S.~D., {Schaye}, J., {Walth}, G.~L., {et~al.} 2022, \apjl, 940, L40, \dodoi{10.3847/2041-8213/aca28e}

\bibitem[{{Johnson} {et~al.}(2024){Johnson}, {Liu}, {Li}, {Schaye}, {Greene}, {Cantalupo}, {Rudie}, {Qu}, {Chen}, {Rafelski}, {Muzahid}, {Chen}, {Contini}, {Kollatschny}, {Mishra}, {Petitjean}, {Rauch}, \& {Zahedy}}]{Johnson24}
{Johnson}, S.~D., {Liu}, Z.~W., {Li}, J. I.~H., {et~al.} 2024, \apj, 966, 218, \dodoi{10.3847/1538-4357/ad3911}

\bibitem[{{Jun} {et~al.}(2017){Jun}, {Im}, {Kim}, \& {Stern}}]{Jun17}
{Jun}, H.~D., {Im}, M., {Kim}, D., \& {Stern}, D. 2017, \apj, 838, 41, \dodoi{10.3847/1538-4357/aa63f9}

\bibitem[{{Jun} {et~al.}(2020){Jun}, {Assef}, {Bauer}, {Blain}, {D{\'\i}az-Santos}, {Eisenhardt}, {Stern}, {Tsai}, {Wright}, \& {Wu}}]{Jun20}
{Jun}, H.~D., {Assef}, R.~J., {Bauer}, F.~E., {et~al.} 2020, \apj, 888, 110, \dodoi{10.3847/1538-4357/ab5e7b}

\bibitem[{{Kauffmann} {et~al.}(2003)}]{kauf03a}
{Kauffmann}, G., {et~al.} 2003, \mnras, 346, 1055, \dodoi{10.1111/j.1365-2966.2003.07154.x}

\bibitem[{{Kennicutt}(1998)}]{kenn98}
{Kennicutt}, Jr., R.~C. 1998, \araa, 36, 189, \dodoi{10.1146/annurev.astro.36.1.189}

\bibitem[{{Kere{\v s}} {et~al.}(2009){Kere{\v s}}, {Katz}, {Fardal}, {Dav{\'e}}, \& {Weinberg}}]{keres09}
{Kere{\v s}}, D., {Katz}, N., {Fardal}, M., {Dav{\'e}}, R., \& {Weinberg}, D.~H. 2009, \mnras, 395, 160, \dodoi{10.1111/j.1365-2966.2009.14541.x}

\bibitem[{{Kere{\v s}} {et~al.}(2005){Kere{\v s}}, {Katz}, {Weinberg}, \& {Dav{\'e}}}]{kere05}
{Kere{\v s}}, D., {Katz}, N., {Weinberg}, D.~H., \& {Dav{\'e}}, R. 2005, \mnras, 363, 2, \dodoi{10.1111/j.1365-2966.2005.09451.x}

\bibitem[{{Kewley} {et~al.}(2001){Kewley}, {Dopita}, {Sutherland}, {Heisler}, \& {Trevena}}]{kewl01}
{Kewley}, L.~J., {Dopita}, M.~A., {Sutherland}, R.~S., {Heisler}, C.~A., \& {Trevena}, J. 2001, \apj, 556, 121, \dodoi{10.1086/321545}

\bibitem[{{Kewley} {et~al.}(2013){Kewley}, {Maier}, {Yabe}, {Ohta}, {Akiyama}, {Dopita}, \& {Yuan}}]{Kewley13b}
{Kewley}, L.~J., {Maier}, C., {Yabe}, K., {et~al.} 2013, \apjl, 774, L10, \dodoi{10.1088/2041-8205/774/1/L10}

\bibitem[{{Kormendy} \& {Ho}(2013)}]{korm13}
{Kormendy}, J., \& {Ho}, L.~C. 2013, \araa, 51, 511, \dodoi{10.1146/annurev-astro-082708-101811}

\bibitem[{{Lamperti} {et~al.}(2024){Lamperti}, {Arribas}, {Perna}, {Rodr{\'\i}guez Del Pino}, {Circosta}, {P{\'e}rez-Gonz{\'a}lez}, {Bunker}, {Carniani}, {Charlot}, {D'Eugenio}, {Maiolino}, {{\"U}bler}, {Willott}, {Bertola}, {B{\"o}ker}, {Cresci}, {Curti}, {Jones}, {Kumari}, {Parlanti}, {Scholtz}, \& {Venturi}}]{Lamperti24}
{Lamperti}, I., {Arribas}, S., {Perna}, M., {et~al.} 2024, arXiv e-prints, arXiv:2406.10348, \dodoi{10.48550/arXiv.2406.10348}

\bibitem[{{Langen} {et~al.}(2023){Langen}, {Cantalupo}, {Steidel}, {Chen}, {Pezzulli}, \& {Gallego}}]{Langen23}
{Langen}, V., {Cantalupo}, S., {Steidel}, C.~C., {et~al.} 2023, \mnras, 519, 5099, \dodoi{10.1093/mnras/stac3205}

\bibitem[{{Lau} {et~al.}(2016){Lau}, {Prochaska}, \& {Hennawi}}]{Lau16}
{Lau}, M.~W., {Prochaska}, J.~X., \& {Hennawi}, J.~F. 2016, \apjs, 226, 25, \dodoi{10.3847/0067-0049/226/2/25}

\bibitem[{{Luridiana} {et~al.}(2015){Luridiana}, {Morisset}, \& {Shaw}}]{Luridiana15}
{Luridiana}, V., {Morisset}, C., \& {Shaw}, R.~A. 2015, Astron. \& Astrop., 573, A42, \dodoi{10.1051/0004-6361/201323152}

\bibitem[{{Ma} {et~al.}(2024){Ma}, {Goulding}, {Greene}, {Zakamska}, {Wylezalek}, \& {Jiang}}]{Ma24}
{Ma}, Y., {Goulding}, A., {Greene}, J.~E., {et~al.} 2024, \apj, 974, 225, \dodoi{10.3847/1538-4357/ad710c}

\bibitem[{{Marshall} {et~al.}(2023){Marshall}, {Perna}, {Willott}, {Maiolino}, {Scholtz}, {{\"U}bler}, {Carniani}, {Arribas}, {L{\"u}tzgendorf}, {Bunker}, {Charlot}, {Ferruit}, {Jakobsen}, {Rix}, {Rodr{\'\i}guez Del Pino}, {B{\"o}ker}, {Cameron}, {Cresci}, {Curtis-Lake}, {Jones}, {Kumari}, {P{\'e}rez-Gonz{\'a}lez}, \& {Reed}}]{Marshall23}
{Marshall}, M.~A., {Perna}, M., {Willott}, C.~J., {et~al.} 2023, \aap, 678, A191, \dodoi{10.1051/0004-6361/202346113}

\bibitem[{{Mercedes-Feliz} {et~al.}(2023){Mercedes-Feliz}, {Angl{\'e}s-Alc{\'a}zar}, {Hayward}, {Cochrane}, {Terrazas}, {Wellons}, {Richings}, {Faucher-Gigu{\`e}re}, {Moreno}, {Su}, {Hopkins}, {Quataert}, \& {Kere{\v{s}}}}]{Mercedes-Feliz23}
{Mercedes-Feliz}, J., {Angl{\'e}s-Alc{\'a}zar}, D., {Hayward}, C.~C., {et~al.} 2023, \mnras, 524, 3446, \dodoi{10.1093/mnras/stad2079}

\bibitem[{{Momose} {et~al.}(2019){Momose}, {Goto}, {Utsumi}, {Hashimoto}, {Chiang}, {Kim}, {Kashikawa}, {Shimasaku}, \& {Miyazaki}}]{Momose19}
{Momose}, R., {Goto}, T., {Utsumi}, Y., {et~al.} 2019, \mnras, 488, 120, \dodoi{10.1093/mnras/stz1707}

\bibitem[{{Mortazavi} \& {Lotz}(2019)}]{Mortazavi19}
{Mortazavi}, S.~A., \& {Lotz}, J.~M. 2019, \mnras, 487, 1551, \dodoi{10.1093/mnras/stz1331}

\bibitem[{{Murray} {et~al.}(2005){Murray}, {Quataert}, \& {Thompson}}]{murr05}
{Murray}, N., {Quataert}, E., \& {Thompson}, T.~A. 2005, \apj, 618, 569, \dodoi{10.1086/426067}

\bibitem[{{Osterbrock} \& {Ferland}(2006)}]{oste06}
{Osterbrock}, D.~E., \& {Ferland}, G.~J. 2006, {Astrophysics of gaseous nebulae and active galactic nuclei} (Sausalito, CA: University Science Books)

\bibitem[{{O'Sullivan} {et~al.}(2020){O'Sullivan}, {Martin}, {Matuszewski}, {Hoadley}, {Hamden}, {Neill}, {Lin}, \& {Parihar}}]{OSullivan20}
{O'Sullivan}, D.~B., {Martin}, C., {Matuszewski}, M., {et~al.} 2020, \apj, 894, 3, \dodoi{10.3847/1538-4357/ab838c}

\bibitem[{{Parlanti} {et~al.}(2024){Parlanti}, {Carniani}, {{\"U}bler}, {Venturi}, {Circosta}, {D'Eugenio}, {Arribas}, {Bunker}, {Charlot}, {L{\"u}tzgendorf}, {Maiolino}, {Perna}, {Rodr{\'\i}guez Del Pino}, {Willott}, {B{\"o}ker}, {Cameron}, {Chevallard}, {Cresci}, {Jones}, {Kumari}, {Lamperti}, \& {Scholtz}}]{Parlanti24}
{Parlanti}, E., {Carniani}, S., {{\"U}bler}, H., {et~al.} 2024, \aap, 684, A24, \dodoi{10.1051/0004-6361/202347914}

\bibitem[{{Perna} {et~al.}(2023){Perna}, {Arribas}, {Marshall}, {D'Eugenio}, {{\"U}bler}, {Bunker}, {Charlot}, {Carniani}, {Jakobsen}, {Maiolino}, {Rodr{\'\i}guez Del Pino}, {Willott}, {B{\"o}ker}, {Circosta}, {Cresci}, {Curti}, {Husemann}, {Kumari}, {Lamperti}, {P{\'e}rez-Gonz{\'a}lez}, \& {Scholtz}}]{Perna23}
{Perna}, M., {Arribas}, S., {Marshall}, M., {et~al.} 2023, \aap, 679, A89, \dodoi{10.1051/0004-6361/202346649}

\bibitem[{{Perrotta} {et~al.}(2019){Perrotta}, {Hamann}, {Zakamska}, {Alexand roff}, {Rupke}, \& {Wylezalek}}]{perr19}
{Perrotta}, S., {Hamann}, F., {Zakamska}, N.~L., {et~al.} 2019, \mnras, 488, 4126, \dodoi{10.1093/mnras/stz1993}

\bibitem[{{Phillips} \& {Cuesta}(1998)}]{Phillips98}
{Phillips}, J.~P., \& {Cuesta}, L. 1998, \aaps, 133, 381, \dodoi{10.1051/aas:1998328}

\bibitem[{{Planck Collaboration} {et~al.}(2014){Planck Collaboration}, {Ade}, {Aghanim}, {Armitage-Caplan}, {Arnaud}, {Ashdown}, {Atrio-Barandela}, {Aumont}, {Baccigalupi}, {Banday}, \& et~al.}]{Planck13}
{Planck Collaboration}, {Ade}, P.~A.~R., {Aghanim}, N., {et~al.} 2014, Astron. $\&$ Astrop., 571, A16, \dodoi{10.1051/0004-6361/201321591}

\bibitem[{{Prochaska} {et~al.}(2014){Prochaska}, {Lau}, \& {Hennawi}}]{Prochaska14}
{Prochaska}, J.~X., {Lau}, M.~W., \& {Hennawi}, J.~F. 2014, \apj, 796, 140, \dodoi{10.1088/0004-637X/796/2/140}

\bibitem[{{Prochaska} {et~al.}(2011){Prochaska}, {Weiner}, {Chen}, {Mulchaey}, \& {Cooksey}}]{Prochaska11b}
{Prochaska}, J.~X., {Weiner}, B., {Chen}, H.~W., {Mulchaey}, J., \& {Cooksey}, K. 2011, \apj, 740, 91, \dodoi{10.1088/0004-637X/740/2/91}

\bibitem[{{Ramesh} {et~al.}(2023){Ramesh}, {Nelson}, \& {Pillepich}}]{Ramesh23}
{Ramesh}, R., {Nelson}, D., \& {Pillepich}, A. 2023, \mnras, 518, 5754, \dodoi{10.1093/mnras/stac3524}

\bibitem[{{Rich} {et~al.}(2010){Rich}, {Dopita}, {Kewley}, \& {Rupke}}]{rich10}
{Rich}, J.~A., {Dopita}, M.~A., {Kewley}, L.~J., \& {Rupke}, D.~S.~N. 2010, \apj, 721, 505, \dodoi{10.1088/0004-637X/721/1/505}

\bibitem[{{Rich} {et~al.}(2011){Rich}, {Kewley}, \& {Dopita}}]{rich11}
{Rich}, J.~A., {Kewley}, L.~J., \& {Dopita}, M.~A. 2011, \apj, 734, 87, \dodoi{10.1088/0004-637X/734/2/87}

\bibitem[{{Rigby} {et~al.}(2023){Rigby}, {Perrin}, {McElwain}, {Kimble}, {Friedman}, {Lallo}, {Doyon}, {Feinberg}, {Ferruit}, {Glasse}, {Rieke}, {Rieke}, {Wright}, {Willott}, {Colon}, {Milam}, {Neff}, {Stark}, {Valenti}, {Abell}, {Abney}, {Abul-Huda}, {Acton}, {Adams}, {Adler}, {Aguilar}, {Ahmed}, {Albert}, {Alberts}, {Aldridge}, {Allen}, {Altenburg}, {{\'A}lvarez-M{\'a}rquez}, {Alves de Oliveira}, {Andersen}, {Anderson}, {Anderson}, {Argyriou}, {Armstrong}, {Arribas}, {Artigau}, {Arvai}, {Atkinson}, {Bacon}, {Bair}, {Banks}, {Barrientes}, {Barringer}, {Bartosik}, {Bast}, {Baudoz}, {Beatty}, {Bechtold}, {Beck}, {Bergeron}, {Bergkoetter}, {Bhatawdekar}, {Birkmann}, {Blazek}, {Blome}, {Boccaletti}, {B{\"o}ker}, {Boia}, {Bonaventura}, {Bond}, {Bosley}, {Boucarut}, {Bourque}, {Bouwman}, {Bower}, {Bowers}, {Boyer}, {Bradley}, {Brady}, {Braun}, {Breda}, {Bresnahan}, {Bright}, {Britt}, {Bromenschenkel}, {Brooks}, {Brooks}, {Brown}, {Brown}, {Brown}, {Bunker}, {Burger}, {Bushouse}, {Cale}, {Cameron}, {Cameron},
  {Canipe}, {Caplinger}, {Caputo}, {Cara}, {Carey}, {Carniani}, {Carrasquilla}, {Carruthers}, {Case}, {Catherine}, {Chance}, {Chapman}, {Charlot}, {Charlow}, {Chayer}, {Chen}, {Cherinka}, {Chichester}, {Chilton}, {Chonis}, {Clampin}, {Clark}, {Clark}, {Coe}, {Coleman}, {Comber}, {Comeau}, {Connolly}, {Cooper}, {Cooper}, {Coppock}, {Correnti}, {Cossou}, {Coulais}, {Coyle}, {Cracraft}, {Curti}, {Cuturic}, {Davis}, {Davis}, {Dean}, {DeLisa}, {deMeester}, {Dencheva}, {Dencheva}, {DePasquale}, {Deschenes}, {Hunor Detre}, {Diaz}, {Dicken}, {DiFelice}, {Dillman}, {Dixon}, {Doggett}, {Donaldson}, {Douglas}, {DuPrie}, {Dupuis}, {Durning}, {Easmin}, {Eck}, {Edeani}, {Egami}, {Ehrenwinkler}, {Eisenhamer}, {Eisenhower}, {Elie}, {Elliott}, {Elliott}, {Ellis}, {Engesser}, {Espinoza}, {Etienne}, {Etxaluze}, {Falini}, {Feeney}, {Ferry}, {Filippazzo}, {Fincham}, {Fix}, {Flagey}, {Florian}, {Flynn}, {Fontanella}, {Ford}, {Forshay}, {Fox}, {Franz}, {Fu}, {Fullerton}, {Galkin}, {Galyer}, {Garc{\'\i}a Mar{\'\i}n}, {Gardner},
  {Gardner}, {Garland}, {Garrett}, {Gasman}, {Gaspar}, {Gaudreau}, {Gauthier}, {Geers}, {Geithner}, {Gennaro}, {Giardino}, {Girard}, {Giuliano}, {Glassmire}, {Glauser}, {Glazer}, {Godfrey}, {Golimowski}, {Gollnitz}, {Gong}, {Gonzaga}, {Gordon}, {Gordon}, {Goudfrooij}, {Greene}, {Greenhouse}, {Grimaldi}, {Groebner}, {Grundy}, {Guillard}, {Gutman}, {Ha}, {Haderlein}, {Hagedorn}, {Hainline}, {Haley}, {Hami}, {Hamilton}, {Hammel}, {Hansen}, {Harkins}, {Harr}, {Hart}, {Hart}, {Hartig}, {Hashimoto}, {Haskins}, {Hathaway}, {Havey}, {Hayden}, {Hecht}, {Heller-Boyer}, {Henriques}, {Henry}, {Hermann}, {Hernandez}, {Hesman}, {Hicks}, {Hilbert}, {Hines}, {Hoffman}, {Holfeltz}, {Holler}, {Hoppa}, {Hott}, {Howard}, {Howard}, {Hunter}, {Hunter}, {Hurst}, {Husemann}, {Hustak}, {Ilinca Ignat}, {Illingworth}, {Irish}, {Jackson}, {Jahromi}, {Jakobsen}, {James}, {James}, {Januszewski}, {Jenkins}, {Jirdeh}, {Johnson}, {Johnson}, {Jones}, {Jones}, {Jones}, {Jones}, {Jordan}, {Jordan}, {Jurczyk}, {Jurling}, {Kaleida}, {Kalmanson},
  {Kammerer}, {Kang}, {Kao}, {Karakla}, {Kavanagh}, {Kelly}, {Kendrew}, {Kennedy}, {Kenny}, {Keski-kuha}, {Keyes}, {Kidwell}, {Kinzel}, {Kirk}, {Kirkpatrick}, {Kirshenblat}, {Klaassen}, {Knapp}, {Knight}, {Knollenberg}, {Koehler}, {Koekemoer}, {Kovacs}, {Kulp}, {Kumari}, {Kyprianou}, {La Massa}, {Labador}, {Labiano}, {Lagage}, {Lajoie}, {Lallo}, {Lam}, {Lamb}, {Lambros}, {Lampenfield}, {Langston}, {Larson}, {Law}, {Lawrence}, {Lee}, {Leisenring}, {Lepo}, {Leveille}, {Levenson}, {Levine}, {Levy}, {Lewis}, {Lewis}, {Libralato}, {Lightsey}, {Link}, {Liu}, {Lo}, {Lockwood}, {Logue}, {Long}, {Long}, {Loomis}, {Lopez-Caniego}, {Lorenzo Alvarez}, {Love-Pruitt}, {Lucy}, {Luetzgendorf}, {Maghami}, {Maiolino}, {Major}, {Malla}, {Malumuth}, {Manjavacas}, {Mannfolk}, {Marrione}, {Marston}, {Martel}, {Maschmann}, {Masci}, {Masciarelli}, {Maszkiewicz}, {Mather}, {McKenzie}, {McLean}, {McMaster}, {Melbourne}, {Mel{\'e}ndez}, {Menzel}, {Merz}, {Meyett}, {Meza}, {Miskey}, {Misselt}, {Moller}, {Morrison}, {Morse}, {Moseley},
  {Mosier}, {Mountain}, {Mueckay}, {Mueller}, {Mullally}, {Murphy}, {Murray}, {Murray}, {Mustelier}, {Muzerolle}, {Mycroft}, {Myers}, {Myrick}, {Nanavati}, {Nance}, {Nayak}, {Naylor}, {Nelan}, {Nickson}, {Nielson}, {Nieto-Santisteban}, {Nikolov}, {Noriega-Crespo}, {O'Shaughnessy}, {O'Sullivan}, {Ochs}, {Ogle}, {Oleszczuk}, {Olmsted}, {Osborne}, {Ottens}, {Owens}, {Pacifici}, {Pagan}, {Page}, {Park}, {Parrish}, {Patapis}, {Paul}, {Pauly}, {Pavlovsky}, {Pedder}, {Peek}, {Pena-Guerrero}, {Penanen}, {Perez}, {Perna}, {Perriello}, {Phillips}, {Pietraszkiewicz}, {Pinaud}, {Pirzkal}, {Pitman}, {Piwowar}, {Platais}, {Player}, {Plesha}, {Pollizi}, {Polster}, {Pontoppidan}, {Porterfield}, {Proffitt}, {Pueyo}, {Pulliam}, {Quirt}, {Quispe Neira}, {Ramos Alarcon}, {Ramsay}, {Rapp}, {Rapp}, {Rauscher}, {Ravindranath}, {Rawle}, {Regan}, {Reichard}, {Reis}, {Ressler}, {Rest}, {Reynolds}, {Rhue}, {Richon}, {Rickman}, {Ridgaway}, {Ritchie}, {Rix}, {Robberto}, {Robinson}, {Robinson}, {Robinson}, {Rock}, {Rodriguez}, {Rodriguez
  Del Pino}, {Roellig}, {Rohrbach}, {Roman}, {Romelfanger}, {Rose}, {Roteliuk}, {Roth}, {Rothwell}, {Rowlands}, {Roy}, {Royer}, {Royle}, {Rui}, {Rumler}, {Runnels}, {Russ}, {Rustamkulov}, {Ryden}, {Ryer}, {Sabata}, {Sabatke}, {Sabbi}, {Samuelson}, {Sapp}, {Sappington}, {Sargent}, {Sauer}, {Scheithauer}, {Schlawin}, {Schlitz}, {Schmitz}, {Schneider}, {Schreiber}, {Schulze}, {Schwab}, {Scott}, {Sembach}, {Shanahan}, {Shaughnessy}, {Shaw}, {Shawger}, {Shay}, {Sheehan}, {Shen}, {Sherman}, {Shiao}, {Shih}, {Shivaei}, {Sienkiewicz}, {Sing}, {Sirianni}, {Sivaramakrishnan}, {Skipper}, {Sloan}, {Slocum}, {Slowinski}, {Smith}, {Smith}, {Smith}, {Smith}, {Snyder}, {Soh}, {Sohn}, {Soto}, {Spencer}, {Stallcup}, {Stansberry}, {Starr}, {Starr}, {Stewart}, {Stiavelli}, {Straughn}, {Strickland}, {Stys}, {Summers}, {Sun}, {Sunnquist}, {Swade}, {Swam}, {Swaters}, {Swoish}, {Taylor}, {Taylor}, {Te Plate}, {Tea}, {Teague}, {Telfer}, {Temim}, {Thatte}, {Thompson}, {Thompson}, {Thomson}, {Tikkanen}, {Tippet}, {Todd}, {Toolan},
  {Tran}, {Trejo}, {Truong}, {Tsukamoto}, {Tustain}, {Tyra}, {Ubeda}, {Underwood}, {Uzzo}, {Van Campen}, {Vandal}, {Vandenbussche}, {Vila}, {Volk}, {Wahlgren}, {Waldman}, {Walker}, {Wander}, {Warfield}, {Warner}, {Wasiak}, {Watkins}, {Weaver}, {Weilert}, {Weiser}, {Weiss}, {Weissman}, {Welty}, {West}, {Wheate}, {Wheatley}, {Wheeler}, {White}, {Whiteaker}, {Whitehouse}, {Whiteleather}, {Whitman}, {Williams}, {Willmer}, {Willoughby}, {Wilson}, {Wirth}, {Wislowski}, {Wolf}, {Wolfe}, {Wolff}, {Workman}, {Wright}, {Wu}, {Wu}, {Wymer}, {Yates}, {Yeager}, {Yeates}, {Yerger}, {Yoon}, {Young}, {Yu}, {Zak}, {Zeidler}, {Zhou}, {Zielinski}, {Zincke}, \& {Zonak}}]{Rigby23}
{Rigby}, J., {Perrin}, M., {McElwain}, M., {et~al.} 2023, \pasp, 135, 048001, \dodoi{10.1088/1538-3873/acb293}

\bibitem[{{Robitaille} {et~al.}(2020){Robitaille}, {Deil}, \& {Ginsburg}}]{Robitaille20}
{Robitaille}, T., {Deil}, C., \& {Ginsburg}, A. 2020, {reproject: Python-based astronomical image reprojection}, Astrophysics Source Code Library, record ascl:2011.023

\bibitem[{Robitaille {et~al.}(2023)Robitaille, Ginsburg, Mumford, Sipőcz, Deil, Kooten, Lim, Biscani, Williams, jimboH, Craig, Stansby, Barentsen, AlistairSymonds, Singer, \& Tollerud}]{thomas_robitaille_2023_7584411}
Robitaille, T., Ginsburg, A., Mumford, S., {et~al.} 2023, astropy/reproject: v0.10.0, v0.10.0,  Zenodo, \dodoi{10.5281/zenodo.7584411}

\bibitem[{{Rudie} {et~al.}(2019){Rudie}, {Steidel}, {Pettini}, {Trainor}, {Strom}, {Hummels}, {Reddy}, \& {Shapley}}]{Rudie19}
{Rudie}, G.~C., {Steidel}, C.~C., {Pettini}, M., {et~al.} 2019, \apj, 885, 61, \dodoi{10.3847/1538-4357/ab4255}

\bibitem[{{Runnoe} {et~al.}(2012){Runnoe}, {Brotherton}, \& {Shang}}]{Runnoe12opt}
{Runnoe}, J.~C., {Brotherton}, M.~S., \& {Shang}, Z. 2012, \mnras, 422, 478, \dodoi{10.1111/j.1365-2966.2012.20620.x}

\bibitem[{{Rupke} {et~al.}(2023{\natexlab{a}}){Rupke}, {Wylezalek}, {Zakamska}, {Veilleux}, {Vayner}, {Bertemes}, {Ishikawa}, {Liu}, {Lim}, {Murphree}, {Whitesell}, {McCrory}, \& {Anicetti}}]{q3dfit23}
{Rupke}, D., {Wylezalek}, D., {Zakamska}, N., {et~al.} 2023{\natexlab{a}}, {q3dfit: PSF decomposition and spectral analysis for JWST-IFU spectroscopy}, Astrophysics Source Code Library, record ascl:2310.004

\bibitem[{{Rupke}(2014)}]{ifsfit2014}
{Rupke}, D. S.~N. 2014, {IFSFIT: Spectral Fitting for Integral Field Spectrographs}.
\newblock \doeprint{1409.005}

\bibitem[{{Rupke} \& {Veilleux}(2015)}]{rupk15}
{Rupke}, D.~S.~N., \& {Veilleux}, S. 2015, \apj, 801, 126, \dodoi{10.1088/0004-637X/801/2/126}

\bibitem[{{Rupke} {et~al.}(2019){Rupke}, {Coil}, {Geach}, {Tremonti}, {Diamond-Stanic}, {George}, {Hickox}, {Kepley}, {Leung}, {Moustakas}, {Rudnick}, \& {Sell}}]{Rupke19}
{Rupke}, D. S.~N., {Coil}, A., {Geach}, J.~E., {et~al.} 2019, \nat, 574, 643, \dodoi{10.1038/s41586-019-1686-1}

\bibitem[{{Rupke} {et~al.}(2023{\natexlab{b}}){Rupke}, {Wylezalek}, {Zakamska}, {Veilleux}, {Bertemes}, {Ishikawa}, {Liu}, {Sankar}, {Vayner}, {Grace Lim}, {McCrory}, {Murphree}, {Whitesell}, {Shen}, {Liu}, {Barrera-Ballesteros}, {Chen}, {Diachenko}, {Goulding}, {Greene}, {Hainline}, {Hamann}, {Heckman}, {Johnson}, {Lutz}, {L{\"u}tzgendorf}, {Mainieri}, {Nesvadba}, {Ogle}, \& {Sturm}}]{Rupke23}
{Rupke}, D. S.~N., {Wylezalek}, D., {Zakamska}, N.~L., {et~al.} 2023{\natexlab{b}}, \apjl, 953, L26, \dodoi{10.3847/2041-8213/aced85}

\bibitem[{{Sabhlok} {et~al.}(2024){Sabhlok}, {Wright}, {Vayner}, {Simonaitis-Boyd}, {Murray}, {Armus}, {Cosens}, {Wiley}, \& {Kriek}}]{Sabhlok24}
{Sabhlok}, S., {Wright}, S.~A., {Vayner}, A., {et~al.} 2024, \apj, 964, 84, \dodoi{10.3847/1538-4357/ad2350}

\bibitem[{{Stern} {et~al.}(2014){Stern}, {Laor}, \& {Baskin}}]{Stern14}
{Stern}, J., {Laor}, A., \& {Baskin}, A. 2014, \mnras, 438, 901, \dodoi{10.1093/mnras/stt1843}

\bibitem[{{Storey} \& {Zeippen}(2000)}]{Storey99}
{Storey}, P.~J., \& {Zeippen}, C.~J. 2000, Mon. Not. R. Astron. S, 312, 813, \dodoi{10.1046/j.1365-8711.2000.03184.x}

\bibitem[{{Strom} {et~al.}(2017){Strom}, {Steidel}, {Rudie}, {Trainor}, {Pettini}, \& {Reddy}}]{Strom17}
{Strom}, A.~L., {Steidel}, C.~C., {Rudie}, G.~C., {et~al.} 2017, \apj, 836, 164, \dodoi{10.3847/1538-4357/836/2/164}

\bibitem[{{Sun} {et~al.}(2024){Sun}, {Fan}, {Han}, {Knudsen}, {Chen}, \& {Zhang}}]{Sun24}
{Sun}, W., {Fan}, L., {Han}, Y., {et~al.} 2024, \apj, 964, 95, \dodoi{10.3847/1538-4357/ad22e3}

\bibitem[{{Temple} {et~al.}(2019){Temple}, {Banerji}, {Hewett}, {Coatman}, {Maddox}, \& {Peroux}}]{Temple19}
{Temple}, M.~J., {Banerji}, M., {Hewett}, P.~C., {et~al.} 2019, \mnras, 487, 2594, \dodoi{10.1093/mnras/stz1420}

\bibitem[{{Thompson} {et~al.}(2015){Thompson}, {Fabian}, {Quataert}, \& {Murray}}]{thom15}
{Thompson}, T.~A., {Fabian}, A.~C., {Quataert}, E., \& {Murray}, N. 2015, \mnras, 449, 147, \dodoi{10.1093/mnras/stv246}

\bibitem[{{Thornton} {et~al.}(1998){Thornton}, {Gaudlitz}, {Janka}, \& {Steinmetz}}]{Thornton98}
{Thornton}, K., {Gaudlitz}, M., {Janka}, H.~T., \& {Steinmetz}, M. 1998, \apj, 500, 95, \dodoi{10.1086/305704}

\bibitem[{{Travascio} {et~al.}(2020){Travascio}, {Zappacosta}, {Cantalupo}, {Piconcelli}, {Arrigoni Battaia}, {Ginolfi}, {Bischetti}, {Vietri}, {Bongiorno}, {D'Odorico}, {Duras}, {Feruglio}, {Vignali}, \& {Fiore}}]{Travascio20}
{Travascio}, A., {Zappacosta}, L., {Cantalupo}, S., {et~al.} 2020, \aap, 635, A157, \dodoi{10.1051/0004-6361/201936197}

\bibitem[{{Tsai} {et~al.}(2018){Tsai}, {Eisenhardt}, {Jun}, {Wu}, {Assef}, {Blain}, {D{\'\i}az-Santos}, {Jones}, {Stern}, {Wright}, \& {Yeh}}]{Tsai18}
{Tsai}, C.-W., {Eisenhardt}, P. R.~M., {Jun}, H.~D., {et~al.} 2018, \apj, 868, 15, \dodoi{10.3847/1538-4357/aae698}

\bibitem[{{Tumlinson} {et~al.}(2017){Tumlinson}, {Peeples}, \& {Werk}}]{Tumlinson17}
{Tumlinson}, J., {Peeples}, M.~S., \& {Werk}, J.~K. 2017, \araa, 55, 389, \dodoi{10.1146/annurev-astro-091916-055240}

\bibitem[{{Tumlinson} {et~al.}(2011){Tumlinson}, {Thom}, {Werk}, {Prochaska}, {Tripp}, {Weinberg}, {Peeples}, {O'Meara}, {Oppenheimer}, {Meiring}, {Katz}, {Dav{\'e}}, {Ford}, \& {Sembach}}]{tuml11}
{Tumlinson}, J., {Thom}, C., {Werk}, J.~K., {et~al.} 2011, Science, 334, 948, \dodoi{10.1126/science.1209840}

\bibitem[{{van de Voort} {et~al.}(2011){van de Voort}, {Schaye}, {Booth}, {Haas}, \& {Dalla Vecchia}}]{vandeVoort11}
{van de Voort}, F., {Schaye}, J., {Booth}, C.~M., {Haas}, M.~R., \& {Dalla Vecchia}, C. 2011, \mnras, 414, 2458, \dodoi{10.1111/j.1365-2966.2011.18565.x}

\bibitem[{{Vayner} {et~al.}(2023{\natexlab{a}}){Vayner}, {Zakamska}, {Sabhlok}, {Wright}, {Armus}, {Murray}, {Walth}, \& {Ishikawa}}]{Vayner23a}
{Vayner}, A., {Zakamska}, N.~L., {Sabhlok}, S., {et~al.} 2023{\natexlab{a}}, \mnras, 519, 961, \dodoi{10.1093/mnras/stac3537}

\bibitem[{{Vayner} {et~al.}(2021){Vayner}, {Wright}, {Murray}, {Armus}, {Boehle}, {Cosens}, {Larkin}, {Mieda}, \& {Walth}}]{Vayner21_ion}
{Vayner}, A., {Wright}, S.~A., {Murray}, N., {et~al.} 2021, \apj, 910, 44, \dodoi{10.3847/1538-4357/abddc1}

\bibitem[{{Vayner} {et~al.}(2023{\natexlab{b}}){Vayner}, {Zakamska}, {Ishikawa}, {Sankar}, {Wylezalek}, {Rupke}, {Veilleux}, {Bertemes}, {Barrera-Ballesteros}, {Chen}, {Diachenko}, {Goulding}, {Greene}, {Hainline}, {Hamann}, {Heckman}, {Johnson}, {Grace Lim}, {Liu}, {Lutz}, {L{\"u}tzgendorf}, {Mainieri}, {McCrory}, {Murphree}, {Nesvadba}, {Ogle}, {Sturm}, \& {Whitesell}}]{Vayner23b}
{Vayner}, A., {Zakamska}, N.~L., {Ishikawa}, Y., {et~al.} 2023{\natexlab{b}}, \apj, 955, 92, \dodoi{10.3847/1538-4357/ace784}

\bibitem[{{Vayner} {et~al.}(2024){Vayner}, {Zakamska}, {Ishikawa}, {Sankar}, {Wylezalek}, {Rupke}, {Veilleux}, {Bertemes}, {Barrera-Ballesteros}, {Chen}, {Diachenko}, {Goulding}, {Greene}, {Hainline}, {Hamann}, {Heckman}, {Johnson}, {Grace Lim}, {Liu}, {Lutz}, {L{\"u}tzgendorf}, {Mainieri}, {McCrory}, {Murphree}, {Nesvadba}, {Ogle}, {Sturm}, \& {Whitesell}}]{Vayner24}
---. 2024, \apj, 960, 126, \dodoi{10.3847/1538-4357/ad0be9}

\bibitem[{{Veilleux} {et~al.}(2023){Veilleux}, {Liu}, {Vayner}, {Wylezalek}, {Rupke}, {Zakamska}, {Ishikawa}, {Bertemes}, {Barrera-Ballesteros}, {Chen}, {Diachenko}, {Goulding}, {Greene}, {Hainline}, {Hamann}, {Heckman}, {Johnson}, {Grace Lim}, {Lutz}, {L{\"u}tzgendorf}, {Mainieri}, {Maiolino}, {McCrory}, {Murphree}, {Nesvadba}, {Ogle}, {Sankar}, {Sturm}, \& {Whitesell}}]{Veilleux23}
{Veilleux}, S., {Liu}, W., {Vayner}, A., {et~al.} 2023, \apj, 953, 56, \dodoi{10.3847/1538-4357/ace10f}

\bibitem[{{Wang} {et~al.}(2024){Wang}, {Wylezalek}, {De Breuck}, {Vernet}, {Rupke}, {Zakamska}, {Vayner}, {Lehnert}, {Nesvadba}, \& {Stern}}]{Wang24}
{Wang}, W., {Wylezalek}, D., {De Breuck}, C., {et~al.} 2024, \aap, 683, A169, \dodoi{10.1051/0004-6361/202348531}

\bibitem[{{Wellons} {et~al.}(2023){Wellons}, {Faucher-Gigu{\`e}re}, {Hopkins}, {Quataert}, {Angl{\'e}s-Alc{\'a}zar}, {Feldmann}, {Hayward}, {Kere{\v{s}}}, {Su}, \& {Wetzel}}]{Wellons23}
{Wellons}, S., {Faucher-Gigu{\`e}re}, C.-A., {Hopkins}, P.~F., {et~al.} 2023, \mnras, 520, 5394, \dodoi{10.1093/mnras/stad511}

\bibitem[{{Wu} {et~al.}(2012){Wu}, {Tsai}, {Sayers}, {Benford}, {Bridge}, {Blain}, {Eisenhardt}, {Stern}, {Petty}, {Assef}, {Bussmann}, {Comerford}, {Cutri}, {Evans}, {Griffith}, {Jarrett}, {Lake}, {Lonsdale}, {Rho}, {Stanford}, {Weiner}, {Wright}, \& {Yan}}]{Wu12}
{Wu}, J., {Tsai}, C.-W., {Sayers}, J., {et~al.} 2012, \apj, 756, 96, \dodoi{10.1088/0004-637X/756/1/96}

\bibitem[{{Wylezalek} {et~al.}(2022){Wylezalek}, {Vayner}, {Rupke}, {Zakamska}, {Veilleux}, {Ishikawa}, {Bertemes}, {Liu}, {Barrera-Ballesteros}, {Chen}, {Goulding}, {Greene}, {Hainline}, {Hamann}, {Heckman}, {Johnson}, {Lutz}, {L{\"u}tzgendorf}, {Mainieri}, {Maiolino}, {Nesvadba}, {Ogle}, \& {Sturm}}]{Wylezalek22}
{Wylezalek}, D., {Vayner}, A., {Rupke}, D. S.~N., {et~al.} 2022, \apjl, 940, L7, \dodoi{10.3847/2041-8213/ac98c3}

\bibitem[{{Zewdie} {et~al.}(2023){Zewdie}, {Assef}, {Mazzucchelli}, {Aravena}, {Blain}, {D{\'\i}az-Santos}, {Eisenhardt}, {Jun}, {Stern}, {Tsai}, \& {Wu}}]{Zewdie23}
{Zewdie}, D., {Assef}, R.~J., {Mazzucchelli}, C., {et~al.} 2023, \aap, 677, A54, \dodoi{10.1051/0004-6361/202346695}

\bibitem[{{Zewdie} {et~al.}(2024){Zewdie}, {Assef}, {Lambert}, {Mazzucchelli}, {Ilani Loubser}, {Aravena}, {Gonz{\'a}lez-L{\'o}pez}, {Jun}, {Tsai}, {Stern}, {Li}, {Fern{\'a}ndez Aranda}, {D{\'\i}az-Santos}, {Eisenhardt}, {Vayner}, {Martin}, {Blain}, \& {Wu}}]{Zewdie24}
{Zewdie}, D., {Assef}, R.~J., {Lambert}, T., {et~al.} 2024, arXiv e-prints, arXiv:2412.04436, \dodoi{10.48550/arXiv.2412.04436}

\bibitem[{{Zhu} \& {M{\'e}nard}(2013{\natexlab{a}})}]{zhu13}
{Zhu}, G., \& {M{\'e}nard}, B. 2013{\natexlab{a}}, \apj, 770, 130, \dodoi{10.1088/0004-637X/770/2/130}

\bibitem[{{Zhu} \& {M{\'e}nard}(2013{\natexlab{b}})}]{Zhu13b}
---. 2013{\natexlab{b}}, \apj, 773, 16, \dodoi{10.1088/0004-637X/773/1/16}

\end{thebibliography}
\bibliographystyle{aasjournal}

\end{document}